\documentclass[aps,prb,twocolumn,superscriptaddress,nofootinbib,
               longbibliography,floatfix]{revtex4-2}

\usepackage{graphicx}%
\usepackage{amsmath,amssymb,amsfonts}%
\usepackage{siunitx}%
\usepackage{bm}%
\usepackage{enumitem}%
\usepackage[breaklinks=true,hidelinks]{hyperref}%

\DeclareSIUnit{\sq}{sq}   

\begin{document}

\title{Emissivity by Design: Geometric Mesh Inductance Governs Thermal Radiation
in Metallic Nanowire Networks}

\author{Amaury Baret}
\email{abaret@uliege.be}
\affiliation{Department of Physics, SPIN, Q-MAT, University of Li\`ege,
All\'ee du Six Ao\^ut 19, B-4000 Li\`ege, Belgium}

\author{Ngoc Duy Nguyen}
\affiliation{Department of Physics, SPIN, Q-MAT, University of Li\`ege,
All\'ee du Six Ao\^ut 19, B-4000 Li\`ege, Belgium}

\begin{abstract}
The thermal infrared emissivity of metallic nanowire networks---the property that
makes them candidate low-emissivity transparent electrodes---follows regularities
that have resisted physical explanation: a strong wire-diameter dependence with no
wire-length dependence, a correlation with sheet resistance that breaks down above
${\sim}\SI{15}{\ohm/\sq}$, and an angular signature that turns from dielectric-like
to metallic as the network densifies. We show that all of them follow from one
local geometric length, the average inter-wire gap $g$, acting through the
\emph{geometric inductance} of a mesh finer than the thermal wavelength:
$X_L = Z_0\,(g/\lambda)\ln(2g/\pi D)$, a parameter-free reactance that exceeds the
ohmic loss and, unlike the DC sheet resistance, stays finite instead of diverging
at the percolation threshold. Since the gap is fixed by areal density and wire
diameter alone, the optics is length-independent and decoupled from percolative
transport; what has been read as a second, optical percolation is instead a smooth
impedance crossing. With a single fitted parameter the theory reproduces 56
silver-nanowire samples across four diameters to a mean absolute error of 0.04,
collapses them onto one universal curve, and---unchanged---predicts the
emissivity--transmittance data of five independent groups and the measured spectral
emissivity, including the silica phonon band. Because this reactance contains no
material constant, the inter-wire gap emerges as the master design variable for the
radiative properties of metallic nanowire transparent conductors.
\end{abstract}


\maketitle

\section{Introduction}\label{sec:intro}
Transparent conductive electrodes based on random metallic nanowire (MNW)
networks now rival indium tin oxide for flexible optoelectronics, transparent
heaters, and energy-efficient glazing~\cite{zhengFirst2026,bardetOptimization2023a,sannicoloMetallic2016a,baretBridge2024b}. For the last two
applications, a property that is decisive yet far less understood than the visible
transmittance or the sheet resistance, is the thermal infrared emissivity
$\varepsilon$: a low $\varepsilon$ turns a transparent electrode into a
``low-e'' coating that suppresses radiative heat exchange, the same function that
makes these films attractive for smart windows and for radiative thermal
management~\cite{hossainRadiative2016,baeTransparent2024,baretLowemissivity2025a,chenSilver2024}. This low-emissivity use of MNW networks is recent,
and the literature on it is growing quickly, but much about it remains poorly
understood. Designing $\varepsilon$ rationally requires knowing what, at the level
of the network microstructure, actually controls it, and that is precisely what
has been missing so far.

Silver nanowires (AgNWs) are by a wide margin the most synthesised and most
characterised member of this family, and they are the only one for which
emissivity has been measured systematically against a controlled microstructure.
Every data set analysed below is therefore an AgNW network. The mechanism we
identify, however, involves no property of silver: it rests on the network
geometry and on the mesh being a good metal, so the picture and its design rules
apply to metallic nanowire networks generally---a point we return to in the
conclusions.

Measurements of $\varepsilon$ on AgNW films have now been reported by several
groups~\cite{hanauerTransparent2021d,pantojaLow2017b,bobingerInfrared2017a,linDirect2019a,bardetOptimization2023a}, and taken together they reveal
a handful of robust and, on their own, puzzling regularities. The most complete
data set to date, reported by Zheng and co-workers and made publicly
available~\cite{zhengFirst2026}, maps $\varepsilon$ against the deposited areal mass density
(the mass of silver per unit film area, denoted $\mathrm{amd}$) for four
well-characterized wire diameters ($D = 58$, $73$, $95$, and $\SI{112}{nm}$) over a
range of angles. Zheng \textit{et~al.} already remarked that the
average inter-wire gap is a key geometric descriptor that tracks the emissivity, but
they did not explain why the gap should be the controlling variable or connect it to
a radiative mechanism~\cite{zhengFirst2026}; supplying that explanation is the purpose of
this work. They also observed that the angular signature changes character
above a well-defined network density, roughly six times the electrical
percolation threshold, and identified this crossover as a second, \emph{optical}
percolation~\cite{zhengFirst2026}. We reproduce that crossover below, and the
phenomenon they report is not in question; what we propose is a different reading
of its origin. In the picture developed here it is not a critical transition of a
connected cluster but the smooth passage of a single sheet impedance through the
free-space scale, driven by the gap---which is why it lies far above the electrical
threshold, and why the ratio between the two densities stays nearly constant with
diameter, as they observe, without either being set by the other.

Experimental results can be formally summarized as follows. The emissivity:
\begin{enumerate}[label=(\roman*),leftmargin=*,topsep=2pt,itemsep=1pt]
\item decreases nonlinearly with areal density, from the bare-substrate value
(${\approx}\,0.9$) toward a metallic minimum (${<}\,0.2$);
\item at fixed density depends strongly on diameter, with \emph{thinner} wires
driving $\varepsilon$ down much faster;
\item is independent of the individual wire \emph{length} at fixed areal density
and diameter;
\item follows a Hagen--Rubens-like $\sqrt{R_s}$ branch against the sheet resistance
$R_s$ at low $R_s$ but deviates sharply above ${\sim}\SI{15}{\ohm/\sq}$, where it
saturates while $R_s$ keeps climbing; and
\item changes angular character with density, from dielectric-like (falling toward
grazing) for sparse networks to metallic (a grazing-angle peak) for dense ones.
\end{enumerate}

\begin{figure*}[!tb]
\centering
\includegraphics[width=\linewidth]{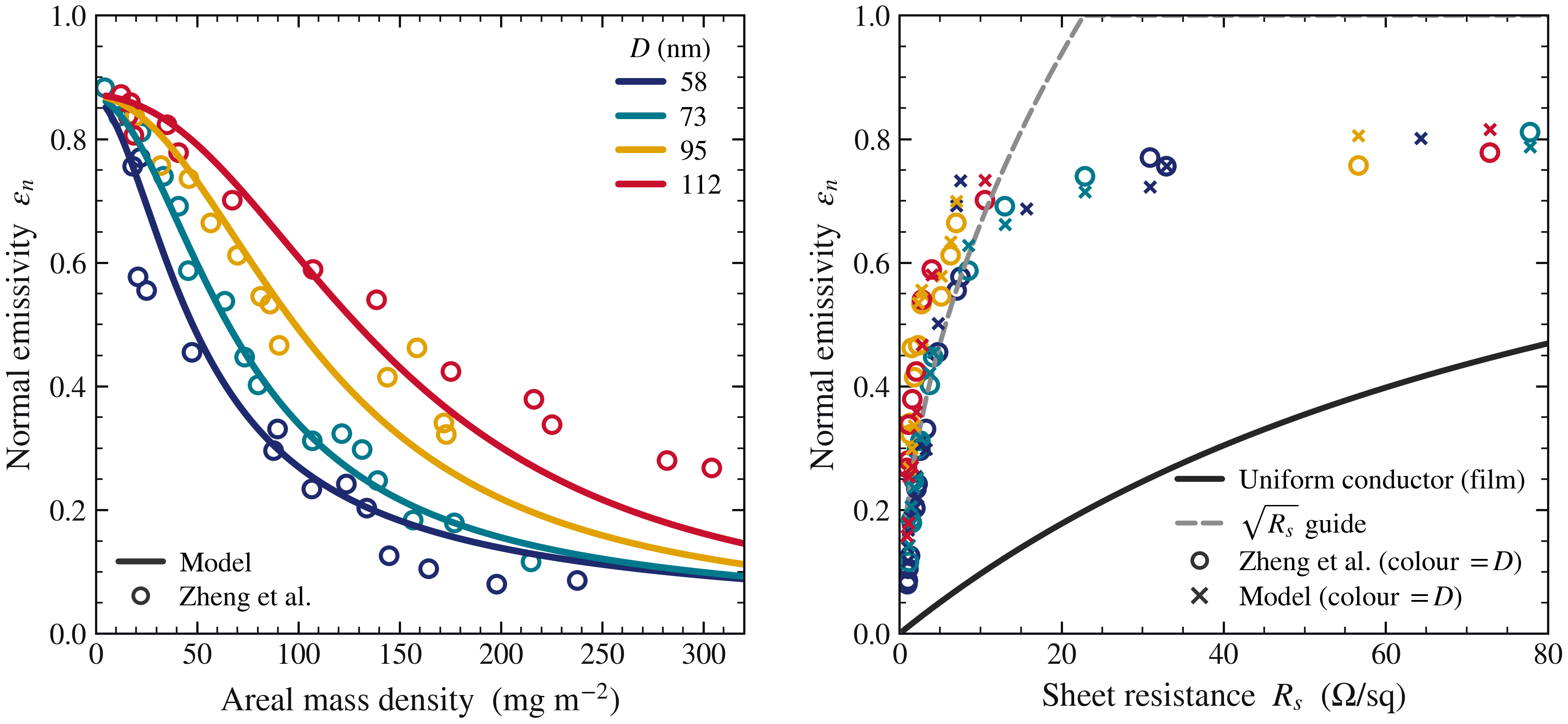}
\caption{The phenomenon and the puzzle. (a) Normal emissivity versus areal mass
density for four diameters; open symbols are the data of Zheng
\textit{et~al.}~\cite{zhengFirst2026} and solid lines the model with one global
parameter set, thinner wires dropping faster as predicted by $g\propto D^2$.
(b) The puzzle: emissivity versus sheet resistance follows a $\sqrt{R_s}$
(Hagen--Rubens) branch at low $R_s$ but deviates above ${\sim}\SI{15}{\ohm/\sq}$,
lying far above the emissivity a uniform conductor of the same $R_s$ would have
(solid black). Crosses are the model prediction (coloured by $D$), placed at each
sample's measured $R_s$, reproducing the same deviation from the uniform-conductor
branch.}
\label{fig:concept}
\end{figure*}

No physical model accounts for this set. Treating the network as a continuous
conductor of resistance $R_s$ and applying the Drude/Hagen--Rubens
relation~\cite{hagenUber1903} fails at once: a uniform conductor with the measured
as-deposited $R_s \approx \SI{5}{\ohm/\sq}$ at
the emissivity mid-point would be an almost perfect mirror
($\varepsilon \approx 0.05$), an order of magnitude below the observed
$\varepsilon \approx 0.45$ [Fig.~\ref{fig:concept}(b)]. Borrowing the network conductivity from percolation or effective-medium transport
theory, which successfully describes the DC behavior of the same
networks~\cite{sannicoloMetallic2016a,pikePercolation1974b,tarasevichElectrical2025}, and feeding it into the same Drude relation is
equally unsatisfactory, because any connectivity-derived conductivity carries a wire-length
dependence that the optics simply does not show. The subwavelength network is of course
still described by a single effective sheet; what fails is identifying that sheet's optical
impedance with a percolative transport conductivity rather than with a local geometric length.
The only published attempt to fit $\varepsilon$ directly, the semi-empirical expression of Hanauer
\textit{et~al.}~\cite{hanauerTransparent2021d}, reproduces the emissivity--transmittance trend but is
purely phenomenological: it contains adjustable
constants without deep microscopic meaning, and its single empirical parameter must be
retuned for every wire geometry---the authors report that it shifts between their three
nanowire batches and leave the dependence unexplained---while already taking an
unphysical value at their own published parameters (Supplementary Materials). It encodes neither the diameter
dependence, the length independence, the angular transition, nor the $R_s$
decoupling.
The community therefore lacks a predictive, physically grounded
picture of what governs infrared emission in these networks. Here we provide one.
We argue that the controlling quantity is neither the sheet resistance nor the
network connectivity but a single local geometric length, the average open gap
between wires $g$; because $g$ is fixed by the areal density and the diameter
alone, the resulting model is parameter-light, explains all five observations,
and collapses the entire body of data onto one universal curve.

\section{Model description}\label{sec:model}
The argument begins with thermodynamics. By Kirchhoff's law~\cite{kirchhoffUeber1860} the
directional spectral emissivity equals the absorptivity, $\varepsilon = A$, and because
the AgNW networks sit on glass that is opaque in the thermal infrared, no light leaves
through the back, so energy conservation gives $\varepsilon = 1 - R$ with $R$ the
front reflectance. Reflection, in turn, is an impedance-mismatch phenomenon: seen
from the incident wave, a surface is a good absorber---and hence, by Kirchhoff, a
good emitter---precisely when its impedance is close to that of free space
($Z_0 = \SI{377}{\ohm}$), and a poor absorber---a good mirror---when its impedance is
near zero.
Determining the emissivity thus reduces to determining the surface impedance that the
thermal-infrared wave sees~\cite{jacksonCLASSICAL1999}. The relevant radiation occupies the
$300\,$K thermal band ($\lambda \approx 5$--$\SI{25}{\micro\meter}$), over which the
integrated emissivity is obtained as a Planck-weighted average (Supplementary
Materials). Throughout this work, we use $\lambda \approx
\SI{10}{\micro\meter}$, near the peak of that band, as a representative wavelength
when quoting numbers. That impedance is simple to characterize because the network
is subwavelength in every structural dimension: the wire diameter
($\sim\SI{100}{nm}$), the film thickness, and the inter-wire gap
($0.3$--$\SI{6}{\micro\meter}$ across the transition that sets $\varepsilon$) all
lie well below $\lambda$. Only a handful of the very sparsest films reach gaps of
$10$--$\SI{26}{\micro\meter}$ and violate this condition, and they sit on the
bare-glass plateau where the exact gap is immaterial; excluding them leaves the
fitted parameter unchanged (Supplementary Materials). The
wave cannot resolve the microstructure; it responds only to the surface current
averaged over the sheet, captured by a single complex sheet admittance $\sigma_s$
(an admittance per square, in units of $\si{\ohm}^{-1}$). Placing
such a sheet at the air/glass interface modifies the Fresnel reflection in the
standard way~\cite{senior,hansonDyadic2008}: the sheet adds a current in parallel with the interface,
and how strongly it shorts the incident field is set by comparing its admittance to
that of free space.
The natural dimensionless measure of that comparison is
$y = Z_0\sigma_s$ (the sheet admittance $\sigma_s$ divided by the free-space
admittance $1/Z_0$, i.e.\ $y = Z_0\sigma_s$), and it controls everything:
$y\!\to\!0$ is a transparent sheet that
recovers the bare-glass emissivity ($\varepsilon \approx 0.86$), $y\!\to\!\infty$ is
a perfectly conducting sheet that shorts the field into a mirror ($\varepsilon \to
0$), and the transition occurs when the sheet impedance $|Z_s| = Z_0/|y|$ falls
through the ${\sim}\SI{100}{\ohm}$ scale. For this air/glass interface the
Planck-band-averaged half-emissivity crossing of the Fresnel model lies at
$|Z_s|\approx\SI{90}{\ohm}$.

The decisive physics is that a wire mesh and a continuous film of the same DC
resistance have very different \emph{optical} impedances. In a film the
alternating current flows in straight sheets; in a mesh it must detour around
every opening, weaving through junctions and enclosing the open cells. A current
that encloses area has inductance, and the larger the cell that the induced
currents must wrap around, the larger that inductance. This geometric, gap-set
inductance presents a reactance $X_L = \omega L$ that, at thermal-IR frequencies,
is comparable to or larger than the ohmic resistance---clearly dominant for the
thicker wires and lower densities, and of the same order as $R_{\mathrm{dc}}$ for
the thinnest wires at high density (Supplementary Materials)---and it is this
reactance, absent in a continuous film, that lifts the optical impedance above the
DC resistance.
The magnetostatics of a strip mesh of period
$g$ and wire width $D$ gives the classic metal-mesh result~\cite{ulrichFarinfrared1967a,marcuvitzWaveguide1986,luukkonenSimple2008}
(see Supplementary Materials for more details)
\begin{equation}
X_L \;=\; Z_0\,\frac{g}{\lambda}\,\ln\!\frac{2g}{\pi D},
\label{eq:XL}
\end{equation}
which carries \emph{no} adjustable parameter; each factor is physical, with
$g/\lambda$ expressing the subwavelength suppression, the logarithm the flux
crowding of thin wires, and $Z_0$ setting the scale. Evaluated at the
mid-transition ($g\approx\SI{1.5}{\micro\meter}$, $D=\SI{58}{nm}$),
Eq.~\eqref{eq:XL} gives $X_L \approx \SI{150}{\ohm}$, the same
${\sim}\SI{100}{\ohm}$ order as the sheet impedance at which the emissivity crosses
one half. The mesh inductance, which the ohmic loss then adds to (Supplementary
Materials), sets this scale---a scale the DC resistance alone cannot supply.

The gap that enters Eq.~\eqref{eq:XL} is not an independent knob; it is fixed by
the synthesis of the material. Conserving mass, the total wire length per unit area is
$\Lambda = 4\,\mathrm{amd}/(\rho_{\mathrm{Ag}}\pi D^2)$, and random lines at line
density $\Lambda$ tile the plane into cells whose mean size, expressed as an
equivalent-circle diameter, is by elementary stochastic geometry
\begin{equation}
g \;=\; k_g\,\frac{2}{\Lambda} \;=\; k_g\,\frac{\rho_{\mathrm{Ag}}\,\pi D^2}{2\,\mathrm{amd}},
\qquad k_g \approx 1.3 .
\label{eq:gap}
\end{equation}
The intermediate steps, from the Poisson line process through the mean cell area to
its equivalent-circle diameter, are derived in the Supplementary Materials.
Two consequences are immediate. First, $g$ depends on the areal density and the
diameter \emph{only}; it is blind to how the total length $\Lambda$ is partitioned
into short or long wires, so any quantity built on $g$ is length-independent by
construction, which is observation~(iii). Second, at fixed areal density
$g \propto D^2$: thicker wires deliver less total length per unit mass, hence
larger gaps, a more inductive mesh, and a higher emissivity, which is
observation~(ii). Two of the five
observations thus fall out of the gap law alone, before any optics is computed.

That the controlling length is $g$, and not the network connectivity, also tells us
what the mechanism is \emph{not}. It is not percolation: the DC giant
cluster appears at a stick threshold $N_c L^2 \approx 5.64$~\cite{liFinitesize2009}, so any
percolation-controlled property carries an explicit wire-length dependence
(through $N_c \propto L^{-2}$) that the optics does not exhibit. Optics is a
\emph{local} response to the gap while transport is a \emph{global} response to
connectivity, and the two need not---and here do not---coincide.

Nor is it a diffraction-grating effect: the gaps ($0.3$--$\SI{6}{\micro\meter}$ over
the transition region) lie below the thermal wavelength and the network is disordered, so the mesh operates
in the quasi-static (long-wavelength) limit---no
diffraction orders propagate and the homogenization into a single $\sigma_s$ is
justified~\cite{ulrichFarinfrared1967a,marcuvitzWaveguide1986,munkFrequency2000,anwarFrequency2018}.

Collecting these ingredients, the sheet admittance of the network is that of a
metallic mesh---a series resistance--inductance branch (the metal) with a small
parallel gap capacitance (the openings),
\begin{equation}
\sigma_s(\omega) \;=\; \frac{1}{R_{\mathrm{dc}}(\omega) - i\,X_L}
\;+\; \big(\!-i\,\omega C_g\big),
\label{eq:sigma}
\end{equation}
with $X_L$ from Eq.~\eqref{eq:XL} and $R_{\mathrm{dc}} =
|\sigma_{\mathrm{Drude}}|^{-1}/(\alpha\,\phi\,D)$ the sheet impedance of the diluted
metal backbone: a full silver film of thickness $D$ would have sheet conductance
$|\sigma_{\mathrm{Drude}}|D$, the mesh covers only a fraction $\phi = \Lambda D$ of
the area with metal, and a further factor $\alpha$ accounts for the part of that
metal that does not carry current efficiently, namely oblique orientation, junction
resistance, and dead ends, so the conductance is reduced by $\alpha\,\phi$ and the
impedance raised by $1/(\alpha\,\phi)$. Because $\omega\tau\approx7$ in this band,
the diluted strip is itself partly reactive through the kinetic inductance of the
carriers, so $R_{\mathrm{dc}}$ is a lumped magnitude for the metal rather than a
purely ohmic term; this affects only the metallic floor at the highest densities and
not the transition region, as quantified in the Supplementary Materials.
The last term, $C_g = \beta\,\varepsilon_0(1+\varepsilon_{\mathrm{sub}})D$, is the
displacement current that fringes across each opening, with the gap acting as a
small capacitor whose plate scale is the wire diameter $D$ and whose dielectric is
air on one side and the substrate on the other; in this band $\omega C_g$ stays below
the metallic branch admittance for the denser films and becomes comparable to it only
for the sparsest networks, where the emissivity is pinned near the bare-glass value
regardless, so this channel does not control the response
(Supplementary Materials) and we retain it only for completeness.
The directional emissivity
then follows from the conducting-sheet Fresnel relations for both polarizations
between air and the (dispersive) glass substrate, $\varepsilon(\theta) =
1 - R(\theta)$, Planck-averaged over the thermal band for the integrated quantity~\cite{modestRadiative2021}.
The model is deliberately minimal. Its leading term, $X_L$, is parameter-free;
the geometric prefactor $k_g$ in Eq.~\eqref{eq:gap} is fixed by the
\emph{independently measured} gap sizes rather than by emissivity; and the gap
capacitance, as noted, does not control the response---removing it entirely (setting
$\beta=0$) changes the global mean absolute error by only $0.004$. This leaves a single
optical fit parameter,
the backbone conduction efficiency $\alpha$---a single lumped factor relating the
connected backbone's sheet conductance to that of an ideal fully aligned film of
the same coverage, absorbing orientation, junction resistance, and dead-end
losses. We fix it at the value that minimises the global mean absolute error,
$\alpha=0.29$, over the entire data set; the choice of error measure is immaterial,
a least-squares fit returning $\alpha=0.27$ and leaving the quoted agreement
unchanged. All material constants are taken from the literature; the result is
robust to them, because $\alpha$ enters only through the product
$\alpha\,\sigma_{\mathrm{dc}}$ and so absorbs factor-of-two changes in the silver
conductivity or relaxation time, and it is likewise insensitive to the choice of
silica dispersion model (Supplementary Materials).

\section{Results and discussion}\label{sec:results}
With the model in hand, we now confront it with the experimental data from literature, beginning with the two
observations that motivated it (Fig.~\ref{fig:concept}). The
emissivity falls from the bare-glass ceiling to a metallic floor as the network
densifies [Fig.~\ref{fig:concept}(a)], and at any fixed density the four diameters
fan out in the order observed, reproduced without diameter-specific tuning.
Figure~\ref{fig:concept}(b) makes the central anomaly quantitative: the data sit
far above the uniform-conductor (Hagen--Rubens) curve and follow it only in trend,
not in magnitude, and the model (crosses, placed at each sample's measured $R_s$)
reproduces this. An AgNW film with $R_s = \SI{1}{\ohm/\sq}$ emits roughly ten
times more than a continuous conductor of the same resistance: its optical sheet
impedance ($|Z_{\mathrm{opt}}|\approx\SI{25}{\ohm}$, a comparable mix of the gap
reactance $X_L\approx\SI{19}{\ohm}$ and the diluted metal branch
$R_{\mathrm{dc}}\approx\SI{17}{\ohm}$) is still far below
$Z_0$ but an order of magnitude above the $\SI{1}{\ohm}$ DC resistance, so the
sheet is a less perfect short and reflects the thermal-IR wave less completely.

\begin{figure*}[!tb]
\centering
\includegraphics[width=\linewidth]{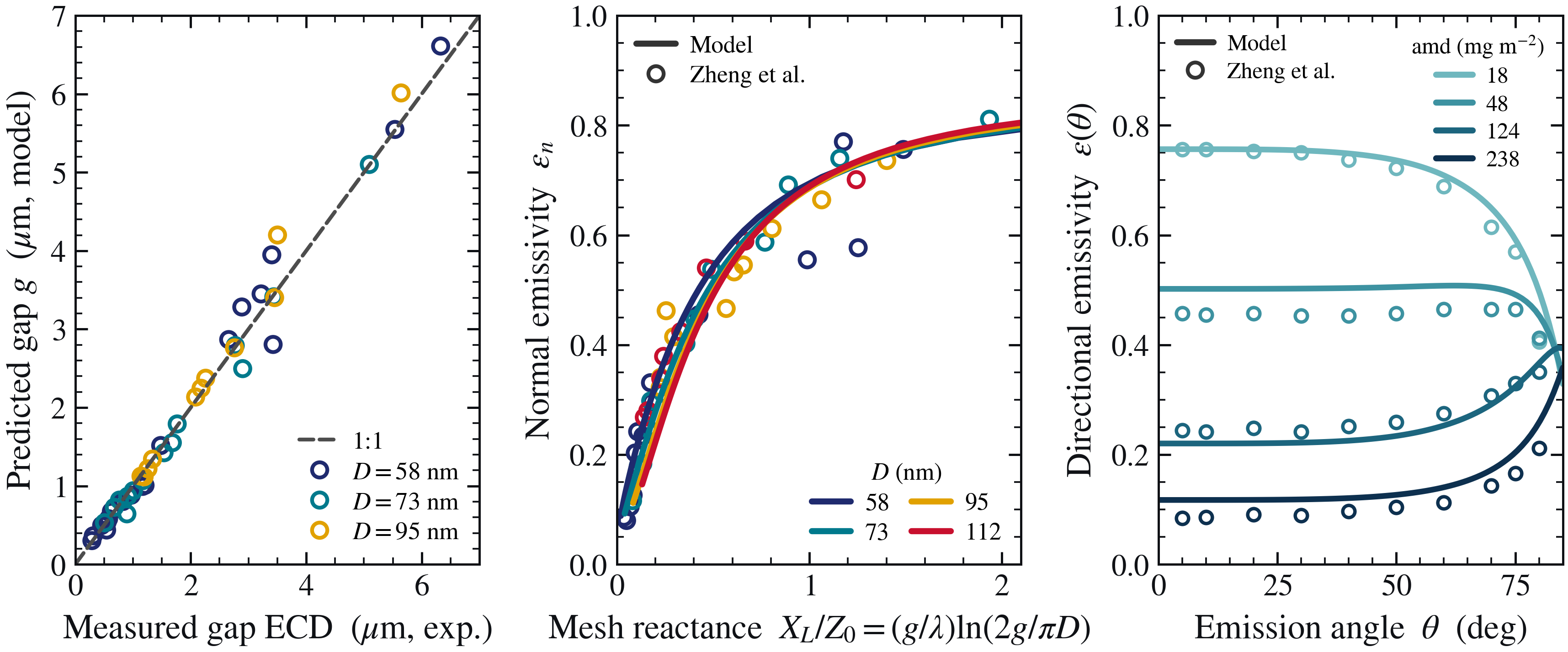}
\caption{The gap is the master variable. (a) The gap predicted from areal density
and diameter [Eq.~\eqref{eq:gap}] against the measured mean gap (equivalent-circle
diameter of the inter-wire voids; see SI) of Zheng \textit{et~al.}~\cite{zhengFirst2026};
a single prefactor $k_g\approx1.3$ collapses the $58,73,95\,$nm diameters onto the
1:1 line. The $\SI{112}{nm}$ gap data are omitted: their gap-area column in the
source file is internally inconsistent (median${>}$mean) and yields a spurious
factor-${\sim}2$ offset, so they are excluded from this geometric test only; the
$\SI{112}{nm}$ emissivity is still reproduced with the same parameters (see SI).
(b) Universal collapse: plotted against the dimensionless mesh reactance
$X_L/Z_0=(g/\lambda)\ln(2g/\pi D)$, the emissivity for all diameters and densities
falls onto one curve. (c) The same sheet admittance reproduces the directional
emissivity across four areal densities ($18,48,124,238\,\mathrm{mg\,m^{-2}}$;
lines, model; open circles, data for $D=\SI{58}{nm}$), tracking the transition
from a dielectric-like fall-off toward grazing (sparse) to a metallic
grazing-angle peak (dense).}
\label{fig:collapse}
\end{figure*}

The geometric law of Eq.~\eqref{eq:gap} is validated directly in
Fig.~\ref{fig:collapse}(a), where the predicted gap matches the measured mean gap from Zheng et al.~\cite{zhengFirst2026}
to a constant prefactor across diameters---an independent test that uses no
emissivity data at all. Its deeper consequence appears in
Fig.~\ref{fig:collapse}(b): when the emissivity is plotted against the single
dimensionless group $X_L/Z_0$, the curves for all four diameters and every density
collapse onto one master curve, and the data collapse with them. The emissivity is
therefore a universal function of the mesh reactance $X_L$---and hence, since $X_L$
is set by $g$ alone, of the gap. Diameter and density are not independent axes but
two routes to the same gap, and once $g$ is fixed the emissivity is fixed with it.
Evaluated versus angle, the same $\sigma_s$ captures the dielectric-to-metallic flip
[Fig.~\ref{fig:collapse}(c)]: a weakly conducting (sparse) sheet reflects like the
bare dielectric and its emissivity falls toward grazing, while a strongly
conducting (dense) sheet shows the metallic grazing-angle peak.

\begin{figure}[!tb]
\centering
\includegraphics[width=\columnwidth]{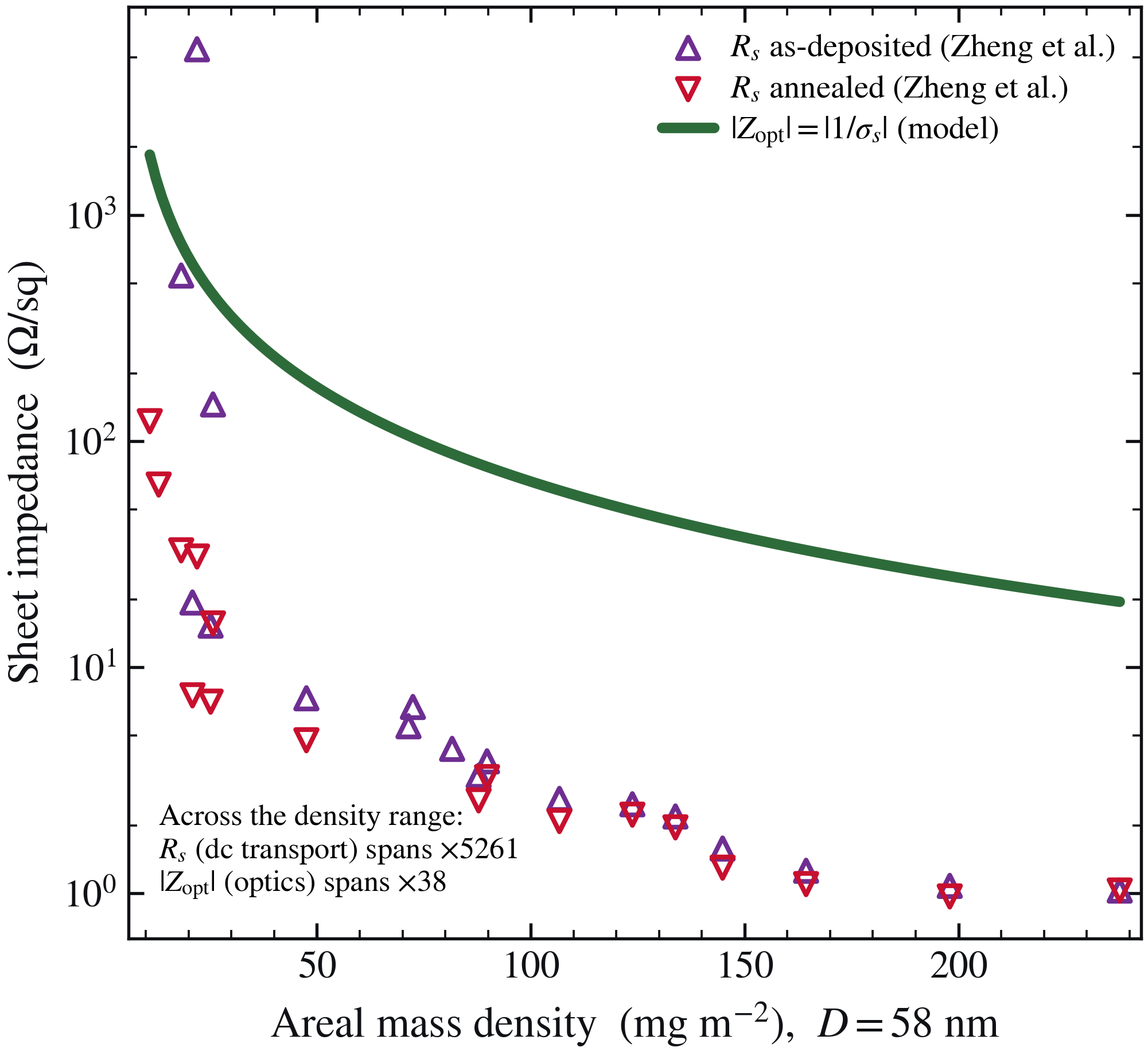}
\caption{Optics and DC transport decouple. Two impedances of the \emph{same}
$D=\SI{58}{nm}$ network are plotted against areal density on common axes: the
\emph{measured} DC sheet resistance $R_s$ (the impedance the DC current sees;
as-deposited, up-triangles; annealed, down-triangles) and the \emph{modeled}
optical sheet impedance $|Z_{\mathrm{opt}}|=|1/\sigma_s|$ that the thermal-IR wave
sees (line). If the emissivity were governed by $R_s$, the two would track; instead
they pull apart. Across the samples for which $R_s$ was measured, densification makes
the as-deposited $R_s$ fall by ${\times}5261$---or ${\times}525$ once the single
most resistive, near-threshold film is set aside---the steep junction/percolation
signature, whereas over that same density range $|Z_{\mathrm{opt}}|$ falls only
${\times}38$. The
emissivity, which the optics sets (Figs.~\ref{fig:concept},\ref{fig:collapse}),
follows the gentle $|Z_{\mathrm{opt}}|$ and not the runaway $R_s$, because optics is
a local response to the gap while transport is a global response to connectivity.
Annealing makes the same point from the other side: fusing the junctions removes the
percolative excess and shifts $R_s$ markedly---its span over these films drops from
${\times}5261$ to ${\times}34$---while the gap, and with it $|Z_{\mathrm{opt}}|$, is
untouched. The two nonetheless remain distinct quantities throughout: even after
annealing, $|Z_{\mathrm{opt}}|$ stays $18$--$81$ times \emph{larger} than $R_s$ at
every density, so the optical impedance is never the sheet resistance the DC current
reports. The complementary length-independence test and the
angular optical threshold are given in the Supplementary Materials.}
\label{fig:decoupling}
\end{figure}

This locality is also what produces observation~(iv), the Hagen--Rubens deviation.
Figure~\ref{fig:decoupling} isolates it by asking the network the same question two
ways: it plots, on common axes, the dc sheet resistance $R_s$ (the impedance felt by
the DC current) and the optical sheet impedance $|Z_{\mathrm{opt}}|=|1/\sigma_s|$
(the impedance felt by the thermal-IR wave) as the film densifies. Were emissivity a
function of $R_s$, the two would coincide; the figure shows instead that they part
company---over the same density range the as-deposited $R_s$ plunges by more than
three orders of magnitude while $|Z_{\mathrm{opt}}|$ changes by
only ${\sim}{\times}38$---because optics and transport are probing different things
about the same network. The thermal-IR
wave drives local oscillating currents that ride the mesh inductance and, only to a
negligible extent, bridge the open gaps as displacement current (the capacitive
channel $\omega C_g$ is dropped throughout, Supplementary Materials); its impedance
is therefore the local metal-branch impedance $R_{\mathrm{dc}}-iX_L$---both terms
set by the finite physical gap and
both staying finite; the DC current, by contrast, must thread a continuous
metallic path and is throttled by the junctions, so $R_s$ diverges toward the
percolation threshold. At high density the two track together---both shrink with
$g$---and the emissivity follows the $\sqrt{R_s}$ branch, but above
${\sim}\SI{15}{\ohm/\sq}$ the as-deposited $R_s$ runs away by orders of magnitude
while $|Z_{\mathrm{opt}}|$ and $\varepsilon$ change only mildly
(Fig.~\ref{fig:decoupling}). The same locality renders the emissivity rigorously
length-independent while $R_s$ remains strongly length-dependent through the
percolation threshold $N_c\propto L^{-2}$ (Supplementary Materials), settling
observation~(iii) and confirming that \textbf{the emissivity is not governed by
connectivity}. The optical crossover of Ref.~\cite{zhengFirst2026} survives this
reading intact---it is reproduced in Fig.~\ref{fig:collapse}(c) and in the
Supplementary Materials---but as a gap-driven impedance crossing rather than a
second percolation: no cluster becomes connected at that density, and
correspondingly the model predicts no length dependence and no critical exponent
there, both of which are testable.

A single global $\alpha$ is fitted once across all four diameters with no
per-diameter tuning, so the diameter fan-out of Fig.~\ref{fig:concept}(a)---which
comes from the geometric law $g\propto D^2$, not from $\alpha$---is already a
prediction rather than a fit. The largest residuals occur at the highest
densities, the regime where the diluted-Drude floor
($R_{\mathrm{dc}}\propto1/\alpha$) matters most and is therefore most sensitive to
the lumped efficiency $\alpha$; there the model mildly underestimates the
thicker-wire emissivity (by ${\sim}0.1$ for $D=95,112\,$nm), consistent with a
single global $\alpha$ that cannot separately tune each diameter's floor. Across
the $56$ samples and four diameters the integrated emissivity is reproduced to a
mean absolute error of $0.04$ ($R^2=0.94$). The same model, with the \emph{same}
$\alpha$, extends beyond this data set with no further tuning: fed only a
calibrated visible transmittance it predicts the emissivity--transmittance data of
five independent groups at wire diameters of $40$--$\SI{120}{nm}$ (mean absolute
error $0.075$ over $23$ points, ranging from $0.02$ to $0.13$ between groups),
and---swapping the constant substrate index for a dispersive fused-silica
model---it reproduces the measured spectral emissivity $\varepsilon(\lambda)$ of
Bardet \textit{et~al.}~\cite{bardetOptimization2023a} from their reported diameter
and areal densities alone, with no quantity adjusted to the spectra. The
$\SI{9}{\micro\meter}$ feature there is the silica reststrahlen band of the
substrate, present already in the bare-glass reference; what the mesh contributes,
and what the model has to get right, is the overall level of each curve and the
$X_L\propto1/\lambda$ slope on which that band is superposed. Both external tests,
and the predicted-versus-measured parity, are given in the Supplementary Materials.

\section{Conclusions}\label{sec:conclusions}
In summary, we have shown that the thermal infrared emissivity of a
metallic-nanowire transparent electrode is governed by a single local geometric
length: the average inter-wire gap. Treating the subwavelength network as one
homogeneous conducting sheet, and recognising that the sheet's optical impedance is
dominated by the geometric inductance of the mesh rather than by its ohmic
resistance, accounts for all five observations~(i)--(v) with one fitted parameter and
collapses heterogeneous data onto a single universal curve. The network is neither a
partial metal film that fills in as wires are added nor a percolating conductor; it
is an inductive mesh whose radiative response is set by the size of its openings.

The consequences for design are direct. Because the gap follows from the areal
density and the wire diameter alone, a target emissivity translates into an explicit
deposition prescription, and thin wires are doubly favorable: at fixed mass they
deliver more total length and tile the plane into smaller cells. Because the optics
is decoupled from percolative transport, that target can be met without chasing the
low sheet resistance conduction would demand---and conversely, a film optimised only
for low $R_s$ need not be a good low-e coating, so the two optima are distinct points
in design space. The length-independence is itself an asset, leaving emissivity
\emph{orthogonal} to the figures of merit that wire length does control through $R_s$
and junction density---flexibility, haze, and processing window---so that each may be
optimised separately, whereas a percolation-controlled emissivity would have locked
them together.

The picture is material- and substrate-agnostic: $X_L$ contains no material
constant, and the metal enters only through $\rho_m$, which sets the gap at a given
areal density, and through its conductivity, which fixes the high-density floor.
What it does require is a mesh that is a good metal, so that $X_L$ remains at least
comparable to the ohmic loss and the two shrink together with the gap. It should
therefore carry over to copper, gold, and hybrid \emph{metallic} nanowire networks
and to metal meshes generally, but not to the far more resistive carbon-nanotube
networks, where $R_{\mathrm{dc}}$ alone would govern the response. This extension
is a prediction: every test reported here is on silver, and a direct measurement on
Cu or Au networks would be the sharpest check of its generality. Three further
tests suggest themselves. Spatially resolved gap statistics, rather than the mean
gap alone, should map onto the spread of $\varepsilon$; a deliberately ordered mesh
of known period would probe $X_L$ without the stochastic-geometry prefactor $k_g$;
and the same $\sigma_s$ validated here on passive films is a natural starting point
for actively tunable low-e surfaces, where a modest electrical, thermal, or
phase-change modulation of the effective gap or substrate permittivity would move
the sheet across the same dielectric-to-metallic transition.

Further details---the derivation of the
gap law and its robustness, the full Fresnel and band-averaging expressions, the
parameter table, the predicted-versus-measured parity and sensitivity analysis, the
external validation against five independent groups and against measured spectral
emissivity, and a quantitative comparison with the Hanauer model---are given in
the Supplementary Materials.

\begin{acknowledgments}
The Authors would like to thank Daniel Bellet for fruitful discussion, and the Authors of Ref.~\cite{zhengFirst2026} for
making their measurements publicly available.
A.B. and N.D.N. acknowledge the financial support from
F.R.S.--FNRS via the CDR project J.0157.24.
\end{acknowledgments}

\section*{Use of generative AI}
We disclose that a generative AI assistant (Anthropic
Claude, Opus 4.8 model) contributed materially to the key conceptual step of this work. The
identification that the empirically observed inter-wire-gap control of the
emissivity is governed by the \emph{geometric inductive reactance} of a
subwavelength metallic mesh---the central idea of the paper---arose from
exploratory reasoning with the AI, which connected two literatures the authors of
the present work had been treating separately: the metal-mesh / frequency-selective-surface
description of periodic conductive grids and the stochastic-geometry and transport
description of random nanowire networks. This cross-domain link is what made the
present model possible. The AI additionally assisted with literature search, with
the implementation and independent cross-checking of the numerical model, and with
language editing. The physical hypothesis so obtained was then tested, derived,
quantified, and validated against experiment by the authors, who verified
every result and take complete responsibility for the content of this article; a
detailed account of the workflow is given in the Supplementary Materials. In
line with publisher policy, the AI is a tool and is not, and cannot be, listed as
an author.

\section*{Declarations}

\noindent\textbf{Conflict of interest.} The authors declare no competing interests.

\smallskip\noindent\textbf{Data availability.} All experimental data analysed in this work are from
published, publicly available sources, which are cited in the text; the most complete
data set is that of Ref.~\cite{zhengFirst2026}, made publicly available by its authors
under a Creative Commons Attribution licence. The specific values used here, together
with a statement of provenance for each, are collected in the code repository below.

\smallskip\noindent\textbf{Code availability.} The complete Python implementation of
the model (gap law, sheet admittance, and Fresnel/Planck averaging), the input data,
and the scripts that regenerate every figure in this article and in the Supplementary
Materials are available under the MIT licence at
\url{https://github.com/abaret-phys/emissivity-by-design}, and from the
corresponding author on request.

\smallskip\noindent\textbf{Author contribution.} A.B. and N.D.N. framed the problem, verified all
derivations and numerical results, and wrote the manuscript. A detailed account of the
workflow, including the role of the generative AI assistant, is given in the
Supplementary Materials.

\bibliography{EmissivityBib}

\end{document}


\title{Supplementary Information for\\
``Emissivity by Design: Geometric Mesh Inductance Governs Thermal
Radiation in Metallic Nanowire Networks''}

\author{Amaury Baret}
\email{abaret@uliege.be}
\affiliation{Department of Physics, SPIN, Q-MAT, University of Li\`ege,
All\'ee du Six Ao\^ut 19, B-4000 Li\`ege, Belgium}

\author{Ngoc Duy Nguyen}
\affiliation{Department of Physics, SPIN, Q-MAT, University of Li\`ege,
All\'ee du Six Ao\^ut 19, B-4000 Li\`ege, Belgium}

\maketitle

\renewcommand{\thefigure}{S\arabic{figure}}
\renewcommand{\thetable}{S\arabic{table}}
\renewcommand{\theequation}{S\arabic{equation}}
\setcounter{figure}{0}
\setcounter{table}{0}
\setcounter{equation}{0}

\section{Overview}

The model of the main text has exactly two ingredients and one adjustable
constant. The first ingredient is a \emph{geometric} law that fixes the mean
inter-wire gap $g$ from the deposition parameters (areal mass density
$\mathrm{amd}$ and wire diameter $D$); the second is the \emph{optical} sheet
admittance $\sigma_s$ of the resulting subwavelength metallic mesh, from which the
emissivity follows through the conducting-sheet Fresnel relations. The single
optical fit parameter is the backbone conduction efficiency $\alpha$.

This document is organised accordingly. Section~\ref{sec:geometry} derives the gap
law and calibrates its one prefactor against independently measured gaps.
Section~\ref{sec:optics} builds the sheet admittance term by term and gives the
Fresnel and band-averaging expressions. Section~\ref{sec:parameters} lists the
complete parameter set. Section~\ref{sec:validation} reports the quantitative
agreement, its robustness, the tests against independent data, and a comparison
with the only prior model. Throughout, ``$\mathrm{dc}$'' quantities are transport
properties and all optical quantities derive from $\sigma_s$; the same global
parameter set is used for every figure.

The Python implementation of everything described below---the gap law, the sheet
admittance, the Fresnel and Planck averaging, and the scripts that regenerate every
figure in the main text and in this document---is available under the MIT licence at
\url{https://github.com/abaret-phys/emissivity-by-design}, together with the input
data and a statement of provenance for every data file.

\section{The gap law from stochastic geometry}
\label{sec:geometry}

\subsection{Mass conservation and the Poisson tessellation}

For wires of diameter $D$ and mass density $\rho_{\mathrm{Ag}}$ deposited to an
areal mass density $\mathrm{amd}$, mass conservation fixes the total wire length
per unit area,
\begin{equation}
\Lambda \;=\; \frac{\mathrm{amd}}{\rho_{\mathrm{Ag}}\,(\pi D^2/4)}
        \;=\; \frac{4\,\mathrm{amd}}{\rho_{\mathrm{Ag}}\,\pi D^2}.
\label{eq:Lambda}
\end{equation}
Equation~\eqref{eq:Lambda} idealises the wire cross-section as a disc of area
$\pi D^2/4$, whereas polyol-grown AgNWs are five-fold twinned and pentagonal. The
error this introduces is a constant factor: for a regular pentagon of
vertex-to-vertex diameter $D$ the cross-sectional area is
$\tfrac{5}{8}\sin(2\pi/5)\,D^2 \approx 0.594\,D^2$ against $\pi D^2/4 \approx
0.785\,D^2$ for the disc, so the circular idealisation overestimates the area---and
hence underestimates $\Lambda$---by ${\approx}32\%$ (it would instead
\emph{under}estimate it by ${\approx}14\%$ if $D$ were read as the flat-to-flat
width, which brackets the true figure between these two conventions). Crucially,
this factor is $D$-independent, so it enters only as a constant rescaling of
$\Lambda$: in the gap law [Eq.~\eqref{eq:gaplaw}] it is exactly degenerate with the
prefactor $k_g$, which we do not compute but calibrate against measured gaps
(Sec.~\ref{sec:calibration}), and in the coverage $\phi=\Lambda D$ it is likewise
absorbed by the fitted efficiency $\alpha$ [Eq.~\eqref{eq:Rdc}], exactly as $\alpha$
absorbs a rescaling of $\sigma_{\mathrm{dc}}$. The cross-section shape therefore
shifts the value of the two fitted constants but no prediction of the model; only a
$D$-dependent departure from a fixed shape would matter, and none is reported over
the $40$--$\SI{120}{nm}$ range considered here.

To turn $\Lambda$ into a typical gap, we model the wires as a stationary, isotropic
Poisson line process of intensity $\Lambda$ (the expected total line length per unit
area), which tiles the plane into convex polygonal cells---the open ``pores'' of the
network. Two standard results of stochastic geometry~\cite{Line2013} give the
typical cell size: the process produces a mean number of cells per unit area equal
to $\Lambda^2/\pi$, so the mean cell area is
\begin{equation}
\langle A\rangle \;=\; \frac{\pi}{\Lambda^2}.
\label{eq:cellarea}
\end{equation}
We express a cell ``size'' as the equivalent-circle diameter (ECD), the diameter of
the disk of equal area, both because it is convenient and because it is how the
measured gaps are reported (below):
\begin{equation}
g_{\mathrm{ECD}} \;=\; 2\sqrt{\frac{\langle A\rangle}{\pi}} \;=\; \frac{2}{\Lambda}.
\label{eq:gecd}
\end{equation}
This is the gap of an \emph{idealised} Poisson network. A real network differs by
order-unity amounts---the wires have finite width rather than being zero-thickness
lines, and the crossings are correlated because junctions are sticky. We absorb
both into a single dimensionless prefactor $k_g$, giving the gap law of the main
text,
\begin{equation}
g \;=\; k_g\,\frac{2}{\Lambda}
  \;=\; k_g\,\frac{\rho_{\mathrm{Ag}}\,\pi D^2}{2\,\mathrm{amd}} .
\label{eq:gaplaw}
\end{equation}

\subsection{Robustness of the scaling}

The Poisson line process is the maximum-entropy (minimal-assumption) model of an
isotropic random set of straight lines and the standard starting point for pore-size
statistics in stochastic fibrous networks~\cite{sampsonSpatial2011,Line2013}. It is
not the only possible description---finite-length-fibre (Mikado/Boolean-segment)
models, correlated-junction models, and other random tessellations have been used
for such networks---but all of them share the same \emph{scaling},
$g\propto 1/\Lambda \propto D^2/\mathrm{amd}$, because that scaling follows from mass
conservation and dimensional analysis alone [Eqs.~\eqref{eq:Lambda}--\eqref{eq:gecd}],
independently of the microscopic tessellation statistics. Only the order-unity
prefactor $k_g$ is model-specific, and we do not compute it from any one
tessellation but fix it directly against measured gaps (below). The choice of the
Poisson process is therefore a convenience for making the $D^2/\mathrm{amd}$ scaling
explicit, not a load-bearing assumption: the two consequences we rely on---length
independence (the gap ignores how $\Lambda$ is partitioned into wires) and the
$g\propto D^2$ diameter ordering---hold for any isotropic random-line description.

\subsection{Experimental gap and calibration of $k_g$}
\label{sec:calibration}

``The gap'' is not a single well-defined length, so we state precisely what is
compared. We do not measure it ourselves: Ref.~\cite{zhengFirst2026} reports, for
every sample, the mean area of the inter-wire voids and the corresponding mean void
ECD, $d_{\mathrm{ECD}} = 2\sqrt{A_{\mathrm{void}}/\pi}$, and we take that tabulated
column directly as the measured gap. This void ECD---rather than any
centre-to-centre or nearest-wire spacing---is the standard descriptor of pore size
in stochastic fibrous networks~\cite{sampsonSpatial2011}, and it is defined exactly
like our model quantity $g_{\mathrm{ECD}}=2/\Lambda$ [Eq.~\eqref{eq:gecd}]. Because
both are ECDs of an opening, a single prefactor $k_g$ aligns them; $k_g$ absorbs the
finite wire width and the junction correlations, not a change of definition.

Fitting $k_g$ to the measured void ECD for $D = 58,73,95\,$nm gives
$k_g \approx 1.3$, with a sample-to-sample scatter of $0.14$ about that value and
per-diameter means spanning only $1.28$--$1.36$; with $k_g$ so fixed, the predicted
and measured gaps satisfy
$g_{\mathrm{pred}}/\mathrm{ECD}_{\mathrm{meas}} = 1.00 \pm 0.10$ across all three
diameters (Fig.~2a, main text). The $112\,$nm gap data are excluded from this calibration:
they yield $g_{\mathrm{pred}}/\mathrm{ECD}_{\mathrm{meas}}\approx 2.0$ (versus
${\approx}1.0$ for the other three diameters), the origin of this discrepancy being
traced to the $112\,$nm gap-area column of the source file being internally
inconsistent---its reported median gap area exceeds its reported mean, which is an
impossible feature for the right-skewed void-area distributions seen at every other
diameter, and hence a corrupted column rather than a breakdown of the geometry law. The omission affects only the $k_g$ calibration;
the $112\,$nm \emph{emissivity} data are retained throughout and reproduced with the
same global parameters (Fig.~1a, main text).

\subsection{Coverage and visible transmittance}

The projected metal coverage follows immediately as $\phi = \Lambda D = 2k_g D/g$. The
visible transmittance, whose extinction scales with coverage, is well described by
$T(550\,\mathrm{nm}) = T_0 - a\,(\mathrm{amd}/D)$ with $T_0 = 0.913$, $a = 0.107$
($R^2 = 0.98$ over $55$ samples). This single empirical relation is the only
ingredient added for the emissivity--transmittance comparison
(Sec.~\ref{sec:validation}); it is independent of the infrared model and does not
enter any other result.

\section{The sheet admittance and the optics}
\label{sec:optics}

The subwavelength network responds as one homogeneous conducting sheet of complex
admittance $\sigma_s(\omega)$, that of a metallic mesh: a series
resistance--inductance metal branch in parallel with a small gap capacitance,
\begin{equation}
\sigma_s(\omega) \;=\; \frac{1}{R_{\mathrm{dc}}(\omega) - i\,X_L}
\;+\; \big(\!-i\,\omega C_g\big)
\qquad\text{(Eq.~3, main text).}
\label{eq:sigma}
\end{equation}
The three terms are the gap-set mesh reactance $X_L$, the diluted-Drude metal term
$R_{\mathrm{dc}}$, and the gap capacitance $C_g$, derived in turn below. The
emissivity then follows from the conducting-sheet Fresnel relations and the
Planck-band average (Sec.~\ref{sec:fresnel}).

\subsection{Mesh reactance $X_L$}

$X_L$ is not a new result: it is the classic shunt susceptance of an inductive grid,
established by Marcuvitz~\cite{marcuvitzWaveguide1986} for a strip grating and by
Ulrich~\cite{ulrichFarinfrared1967a} for a two-dimensional metallic mesh in the far
infrared (the latter also being the transmission-line, shunt-admittance picture we
adopt). Marcuvitz's inductive-grid entry gives the normalized reactance of a grating
of period $a$ and strip width $w$ as
\begin{equation}
\frac{X}{Z_0} \;=\; \frac{a}{\lambda}\,\ln\!\Big(\csc\frac{\pi w}{2a}\Big).
\label{eq:marcuvitz}
\end{equation}
Identifying the period with the inter-wire gap ($a\to g$) and the strip width with
the wire diameter ($w\to D$), the thin-wire limit $D\ll g$ gives
$\csc(\pi D/2g)\to 2g/\pi D$, so Eq.~\eqref{eq:marcuvitz} reduces to the main-text
form
\begin{equation}
X_L \;=\; Z_0\,\frac{g}{\lambda}\,\ln\frac{2g}{\pi D}.
\label{eq:XL}
\end{equation}
Our only step is this passage to the gap $g$; the $\ln(2g/\pi D)$ is the
thin-wire asymptote of the $\ln\csc$ function of
Refs.~\cite{marcuvitzWaveguide1986,ulrichFarinfrared1967a}. The same $\ln\csc$ grid
parameter is re-derived and validated against full-wave simulation by Luukkonen
\textit{et~al.}~\cite{luukkonenSimple2008}, whose inductive strip-grid expression
[their Eq.~(4)] is Eq.~\eqref{eq:marcuvitz} in the present notation---note that
their $D$ denotes the grid \emph{period} and their $w$ the strip width, the inverse
of the convention used here. Strictly, the grating
period is the gap plus the strip width, $a = g+D$, so $a\to g$ is an approximation
rather than a relabelling; it is exact only in the thin-wire limit $D\ll g$ that
also justifies the $\ln\csc$ asymptote. It is excellent over the transition region
that sets the emissivity (where $g$ is a few $\si{\micro\meter}$ and $D\lesssim
\SI{0.1}{\micro\meter}$), and worst at the densest films, where using $a=g+D$ would
raise $X_L$ by ${\approx}37\%$ for the densest $\SI{58}{nm}$ sample and by
${\approx}13$--$21\%$ for the densest $95$ and $\SI{112}{nm}$ ones. Since $k_g$ is
calibrated against the measured gaps, a systematic rescaling of this kind is
absorbed into that prefactor and into $\alpha$; what it does not do is change any
scaling or prediction of the model. In the numerics the
argument of the logarithm is clamped to $\geq 1.05$ as a guard only; it is never
triggered, the smallest value of $2g/\pi D$ across all samples being $3.3$ (the
densest $\SI{58}{nm}$ film). $X_L$ has no parameter to be adjusted.

Equation~\eqref{eq:XL} is a quasi-static result: it presumes $g<\lambda$, so that no
diffracted order propagates and the mesh homogenises into a single sheet. Over the
transition that sets the emissivity the condition is met across essentially the
whole thermal band: there $g\leq\SI{6.6}{\micro\meter}$, so $g<\lambda$ everywhere
above $\SI{6.6}{\micro\meter}$, and the residual violation near the blue edge
carries only ${\approx}6\%$ of the Planck weight at $300\,$K. For the ten sparsest
films, whose gaps run from $6$ to $\SI{26}{\micro\meter}$, the condition fails over a
substantial part of the band and the formula is applied outside its regime. Those
films are all pinned at the bare-glass plateau, where the sheet is too weakly
conducting for its impedance to matter---for the seven with
$g>\SI{10}{\micro\meter}$, even halving the gap moves $\varepsilon$ by no more than
$0.07$---so they carry little information about $X_L$, and
indeed refitting $\alpha$ with every $g>\SI{6}{\micro\meter}$ sample removed returns
the same optimum to four digits ($0.2927$). The quoted agreement is therefore not
propped up by the points where the approximation fails; the transition region, where
the model is tested, lies inside it.

\subsection{The metal term $R_{\mathrm{dc}}$ of the diluted backbone}

The lossy part of the metal branch is built from the bulk material in three
steps. (i) A \emph{continuous} silver film of thickness $D$ has sheet conductance
$\sigma_{\mathrm{Ag}}(\omega)\,D$; we characterise it by the magnitude of the Drude
conductivity, $|\sigma_{\mathrm{Ag}}| =
\sigma_{\mathrm{dc}}/\sqrt{1+(\omega\tau)^2}$. (ii) The network covers only the
projected fraction $\phi = \Lambda D$ of the area with metal (the ``dilution''),
reducing the conductance to $|\sigma_{\mathrm{Ag}}|\,\phi D$. (iii) Not all of that
metal carries current as effectively as an ideal aligned, fully connected film:
wires lie at random angles to the field, junctions add contact resistance, and some
wire length sits in non-load-bearing dead ends. We lump these into a single
dimensionless efficiency $\alpha\in(0,1]$. Collecting the three factors,
\begin{equation}
\frac{1}{R_{\mathrm{dc}}} \;=\; \alpha\,\phi\,|\sigma_{\mathrm{Ag}}(\omega)|\,D
\quad\Longrightarrow\quad
R_{\mathrm{dc}} \;=\; \frac{\sqrt{1+(\omega\tau)^2}}{\alpha\,\phi\,\sigma_{\mathrm{dc}}\,D}.
\label{eq:Rdc}
\end{equation}
Only the product $\alpha\,\sigma_{\mathrm{dc}}$ enters, so $\alpha$ absorbs any
overall rescaling of the silver conductivity---which is why the whole model has
effectively one optical degree of freedom (Sec.~\ref{sec:validation}).

Equation~\eqref{eq:Rdc} deserves a word of interpretation, because at thermal-infrared
frequencies silver is not a purely resistive conductor. With $\tau =
\SI{3.8e-14}{s}$ one has $\omega\tau \approx 7$ at $\lambda =
\SI{10}{\micro\meter}$, so the Drude conductivity $\sigma_{\mathrm{Ag}} =
\sigma_{\mathrm{dc}}/(1-i\omega\tau)$ is dominated by its imaginary part: the
diluted metal strip is itself largely reactive, its \emph{kinetic} inductance
arising from electron inertia rather than from the mesh geometry. Taking the
magnitude $|\sigma_{\mathrm{Ag}}|$ therefore does not isolate the dissipated power,
which is governed by $\mathrm{Re}\,\sigma_{\mathrm{Ag}}$; what it does is preserve
the \emph{magnitude} of the metal-strip sheet impedance,
$|1/(\alpha\phi\sigma_{\mathrm{Ag}}D)|$, which is the quantity that sets how
strongly the sheet shorts the incident field. $R_{\mathrm{dc}}$ as defined in
Eq.~\eqref{eq:Rdc} is thus best read as a lumped magnitude for the metal strip
rather than as a strictly ohmic resistance, and the split between the two terms of
$Z_{\mathrm{metal}} = R_{\mathrm{dc}} - iX_L$ is correspondingly a modelling
convention: part of what Eq.~\eqref{eq:Rdc} carries is kinetic reactance. The
distinction matters only at the highest densities, where the kinetic and geometric
contributions become comparable---for the densest $\SI{58}{nm}$ film the kinetic
reactance reaches ${\approx}\SI{14}{\ohm/\sq}$ against $X_L
\approx\SI{14}{\ohm/\sq}$---so that at the metallic floor the mesh is not the sole
source of the sheet reactance. Everywhere else, and in particular across the
transition region that sets $\varepsilon$ and fixes $\alpha$, the geometric term
$X_L$ dominates the metal branch (the ratios quoted below), and the collapse and
scaling results rest on that regime. Treating the metal branch with the full complex
$\sigma_{\mathrm{Ag}}(\omega)$ instead, so that the kinetic and geometric reactances
add explicitly, leaves the picture intact: the fitted efficiency shifts to
$\alpha\approx0.20$ and the global agreement is essentially unchanged (mean absolute
error $0.048$, $R^2=0.93$).

That $\alpha$ and $\sigma_{\mathrm{dc}}$ enter only as a product matters physically,
because the bulk conductivity is not obviously
appropriate at these diameters. The bulk electron mean free path of silver is
$\ell = v_F\tau \approx \SI{53}{nm}$, comparable to the wire diameters
($D/\ell \approx 1$--$2$ for $D=58$--$\SI{112}{nm}$, and ${\approx}0.8$ for the
thinnest literature wires at $\SI{40}{nm}$). Surface and grain-boundary scattering
therefore shortens the effective mean free path, placing the true single-wire
conductivity \emph{below} bulk $\sigma_{\mathrm{dc}}$ by a Fuchs--Sondheimer factor
of order unity. We do not model this separately: it is folded into $\alpha$.
Consistently, the fitted $\alpha=0.29$ gives an effective conductance
$\alpha\,\sigma_{\mathrm{dc}}\approx1.8\times10^{7}\,\mathrm{S/m}$, about a factor of
three below bulk---the right order for surface-scattering suppression at $D\sim\ell$
combined with the orientation and junction losses that $\alpha$ also lumps in. This
does set a lower validity bound: well below $D\sim\SI{20}{nm}$ the wire ceases to be
a good bulk-like metal, the diluted-Drude branch loses meaning, and the premise that
$X_L$ dominates a still-metallic loss breaks down---the same boundary that excludes
the much more resistive carbon-nanotube networks. Within the $40$--$\SI{112}{nm}$
range of all data considered here the wires remain good metals and the treatment is
justified.

Numerically, at $\lambda = \SI{10}{\micro\meter}$, $X_L$ and $R_{\mathrm{dc}}$ are of
comparable magnitude with $X_L$ the larger: over the measured samples the ratio
$X_L/R_{\mathrm{dc}}$ runs from ${\approx}1.0$ to ${\approx}3.1$ for
$D = \SI{58}{nm}$, rising to ${\approx}5$--$15$ for $D = \SI{112}{nm}$ (largest for
thick wires at low density, order unity for the thinnest wires at high density).
Both $X_L$ and $R_{\mathrm{dc}}$ shrink with $g$, so the universal collapse of
Fig.~2b is preserved; but $R_{\mathrm{dc}}$ is not a small correction---it sets the
high-density floor of $\varepsilon$ and is the reason the model mildly
underestimates the thicker-wire emissivity at the highest densities, where the
lumped $\alpha$ controls the floor most strongly.

\subsection{Circuit topology and the gap capacitance $C_g$}

The three terms enter Eq.~\eqref{eq:sigma} with the topology of a physical mesh.
$R_{\mathrm{dc}}$ and $X_L$ are in \emph{series}: the same surface current that
flows along the metal (dissipating in $R_{\mathrm{dc}}$) is the current that must
detour around the openings and enclose the cells (storing energy in $X_L$), so they
add along one path, $Z_{\mathrm{metal}} = R_{\mathrm{dc}} - iX_L$. The gap
capacitance is in \emph{parallel} with that branch: the displacement current across
the openings is an \emph{alternative} route for the field that bypasses the metal,
and parallel paths add admittances, giving Eq.~\eqref{eq:sigma}. Physically, the two
metal edges bounding a gap, with air above and substrate below, act as a small
fringing capacitor whose plate scale is set by the wire diameter (not the much
larger gap), so
\begin{equation}
C_g \;=\; \beta\,\varepsilon_0\,(1+\varepsilon_{\mathrm{sub}})\,D,
\label{eq:Cg}
\end{equation}
with $\beta$ an order-unity fringe prefactor and $(1+\varepsilon_{\mathrm{sub}})$ the
mean of the air and substrate permittivities. This channel does not control the
emissivity, but not because it is uniformly small: at
$\lambda = \SI{10}{\micro\meter}$, $\omega C_g \approx
4\text{--}7\times10^{-4}\,\mathrm{S/sq}$ (rising with $D$), so it is well below the
metal-branch admittance $1/|R_{\mathrm{dc}}-iX_L|\sim10^{-2}\,\mathrm{S/sq}$ only for
the denser films. For the sparsest samples the metal branch is itself
weak and the ratio $\omega C_g/|1/(R_{\mathrm{dc}}-iX_L)|$ reaches and then exceeds
unity, rising to ${\approx}2.6$ for the most dilute $\SI{73}{nm}$ film and
${\approx}2.9$ for the most dilute $\SI{112}{nm}$ one. The influence of $C_g$
nevertheless remains vanishing, because it is precisely in that dilute regime that
the \emph{total} sheet admittance is small and the emissivity sits at the bare-glass
value $\varepsilon\approx0.86$ regardless of this
term: across the full data set, setting $\beta = 0$ shifts individual predictions by at
most ${\approx}0.05$ (all on high-$\varepsilon$ sparse films) and changes the global
mean absolute error by only $0.004$ ($0.046$ versus $0.042$). We therefore retain
$C_g$ for completeness but the model does not rely on it.

\subsection{Conducting-sheet Fresnel relations and band averaging}
\label{sec:fresnel}

For a sheet of admittance $\sigma_s$ between air ($n_1=1$) and substrate ($n_2$),
the reflection coefficients take the transmission-line/shunt
form~\cite{senior,hansonDyadic2008} with $y_s = Z_0\sigma_s$,
\begin{align}
r_{\mathrm{TE}} &= \frac{Y_1 - Y_2 - y_s}{Y_1 + Y_2 + y_s},
& Y_i &= n_i\cos\theta_i, \\
r_{\mathrm{TM}} &= \frac{Y_1' - Y_2' - y_s}{Y_1' + Y_2' + y_s},
& Y_i' &= n_i/\cos\theta_i,
\end{align}
with $\theta_2$ from Snell's law. The substrate is opaque in the thermal
infrared~\cite{kischkatMidinfrared2012}, so no light escapes through the back and
$\varepsilon(\theta) = 1 - \tfrac{1}{2}(|r_{\mathrm{TE}}|^2 + |r_{\mathrm{TM}}|^2)$.

By ``Planck-averaged'' we mean the spectral emissivity weighted by the Planck
spectral radiance $B_\lambda(T)$ and integrated over the thermal band---the
emitted-power-weighted mean, not a flat or wavenumber-uniform average:
\begin{equation}
\bar{\varepsilon}(\theta) \;=\;
\frac{\displaystyle\int_{\lambda_1}^{\lambda_2} \varepsilon(\theta,\lambda)\,
B_\lambda(T)\,\mathrm{d}\lambda}
{\displaystyle\int_{\lambda_1}^{\lambda_2} B_\lambda(T)\,\mathrm{d}\lambda},
\qquad
B_\lambda(T)=\frac{2hc^2}{\lambda^5}\frac{1}{e^{hc/\lambda k_B T}-1},
\label{eq:planckavg}
\end{equation}
with $[\lambda_1,\lambda_2]=[5,25]\,\si{\micro\meter}$ and
$T=300\,$K~\cite{modestRadiative2021}; the integral is evaluated numerically on the
same wavelength grid used for $\sigma_s$, $41$ points, which is converged: raising
the sampling to $641$ points shifts any individual $\bar\varepsilon$ by at most
$1.6\times10^{-3}$ and the global mean absolute error by less than $10^{-4}$, both
far below the experimental scatter. This is the quantity that governs
radiative heat exchange, weighting each wavelength by how much power the surface
actually radiates there. With the dispersive substrate index $n_2(\lambda)$ of
Sec.~\ref{sec:parameters} the bare film ($\sigma_s\to0$) has Planck-band emissivity
$\varepsilon_{\mathrm{bare}}\approx0.86$, and the half-emissivity crossing of the
air/glass Fresnel model lies at $|Z_s|\approx\SI{90}{\ohm}$---the
${\sim}\SI{100}{\ohm}$ scale quoted in the main text.

\section{Parameters}
\label{sec:parameters}

Table~\ref{tab:params} lists the complete parameter set; the same values are used
for every figure, diameter, density, angle, and wavelength.

\begin{table}[h]
\centering
\caption{Complete parameter set.}
\label{tab:params}
\begin{tabular}{lll}
\toprule
Symbol & Value & Origin \\
\midrule
$\sigma_{\mathrm{dc}}$ (Ag) & $6.3\times10^{7}\,\mathrm{S/m}$ & literature~\cite{ordalOptical1985} \\
$\tau$ (Ag) & $3.8\times10^{-14}\,\mathrm{s}$ & literature~\cite{ordalOptical1985} \\
$\rho_{\mathrm{Ag}}$ & $10490\,\mathrm{kg/m^3}$ & literature~\cite{crcHandbook2016} \\
$n_{\mathrm{glass}}(\lambda)$ & dispersive SiO$_2$ (below) & 3-oscillator Lorentz; $\varepsilon_{\mathrm{bare}}{\approx}0.86$ \\
SiO$_2$ phonons $\omega_{0j}$ & $\{1090, 800, 450\}\,\mathrm{cm^{-1}}$ & bands as in~\cite{kischkatMidinfrared2012} \\
$\alpha$ (backbone efficiency) & $0.29$ & global fit (minimum mean absolute error) \\
$\beta$ (gap capacitance) & $1$ & physical estimate (negligible) \\
$k_g$ (gap prefactor) & $1.3$ & set by measured gap data \\
\bottomrule
\end{tabular}
\end{table}

Of these, only $\alpha$ is adjusted to the emissivity data. It is the single
multiplicative factor relating the connected backbone's sheet conductance
$\alpha\,\phi\,\sigma_{\mathrm{dc}}D$ [Eq.~\eqref{eq:Rdc}] to that of an ideal fully
aligned, fully connected film of the same coverage; it lumps orientation averaging,
junction/contact resistance, and dead-end wire fractions, and we do not decompose
it. We fix it at the value minimising the global mean absolute error,
$\alpha=0.29$, held for the entire data set; minimising the sum of squared
residuals instead returns $\alpha=0.27$ and leaves the agreement of
Sec.~\ref{sec:validation} unchanged (mean absolute error $0.043$ versus $0.042$,
$R^2=0.945$ versus $0.945$), so nothing in what follows depends on which measure is
used. The leading term $X_L$ is parameter-free, $k_g$ is set by the
measured gaps [Eq.~\eqref{eq:gaplaw}], and $\beta$ does not affect the result (setting
it to zero shifts the global MAE by only $0.004$); so $\alpha$ is the
one degree of freedom.

The substrate index is a compact three-oscillator Lorentz model of fused silica,
\begin{equation}
\varepsilon(\omega)=\varepsilon_\infty+\sum_j
\frac{s_j\,\omega_{0j}^2}{\omega_{0j}^2-\omega^2-i\gamma_j\omega},
\qquad n_{\mathrm{glass}}(\lambda)=\sqrt{\varepsilon},
\label{eq:sio2}
\end{equation}
with $\varepsilon_\infty=1.90$, transverse-optical modes at
$\omega_{0j}=\{1090,800,450\}\,\mathrm{cm^{-1}}$ (the Si--O asymmetric stretch,
bending, and rocking bands at ${\approx}\,9$, $12.5$, and $\SI{21}{\micro\meter}$),
oscillator strengths $s_j=\{0.82,0.04,0.66\}$, and dampings
$\gamma_j=\{95,55,45\}\,\mathrm{cm^{-1}}$. This is a representative effective
parametrization---not a fit to our emissivity---whose two upper bands coincide with
the SiO$_2$ reststrahlen and \SIrange{700}{900}{cm^{-1}} features
of~\cite{kischkatMidinfrared2012}. The \emph{same} $n_{\mathrm{glass}}(\lambda)$ is
used for the band-integrated emissivity and for the spectral comparison of
Fig.~\ref{fig:validation}(b), so a single physical substrate underlies every figure.

\section{Validation and robustness}
\label{sec:validation}

\subsection{Overall agreement and sensitivity}

\begin{figure}[h]
\centering
\includegraphics[width=0.6\linewidth]{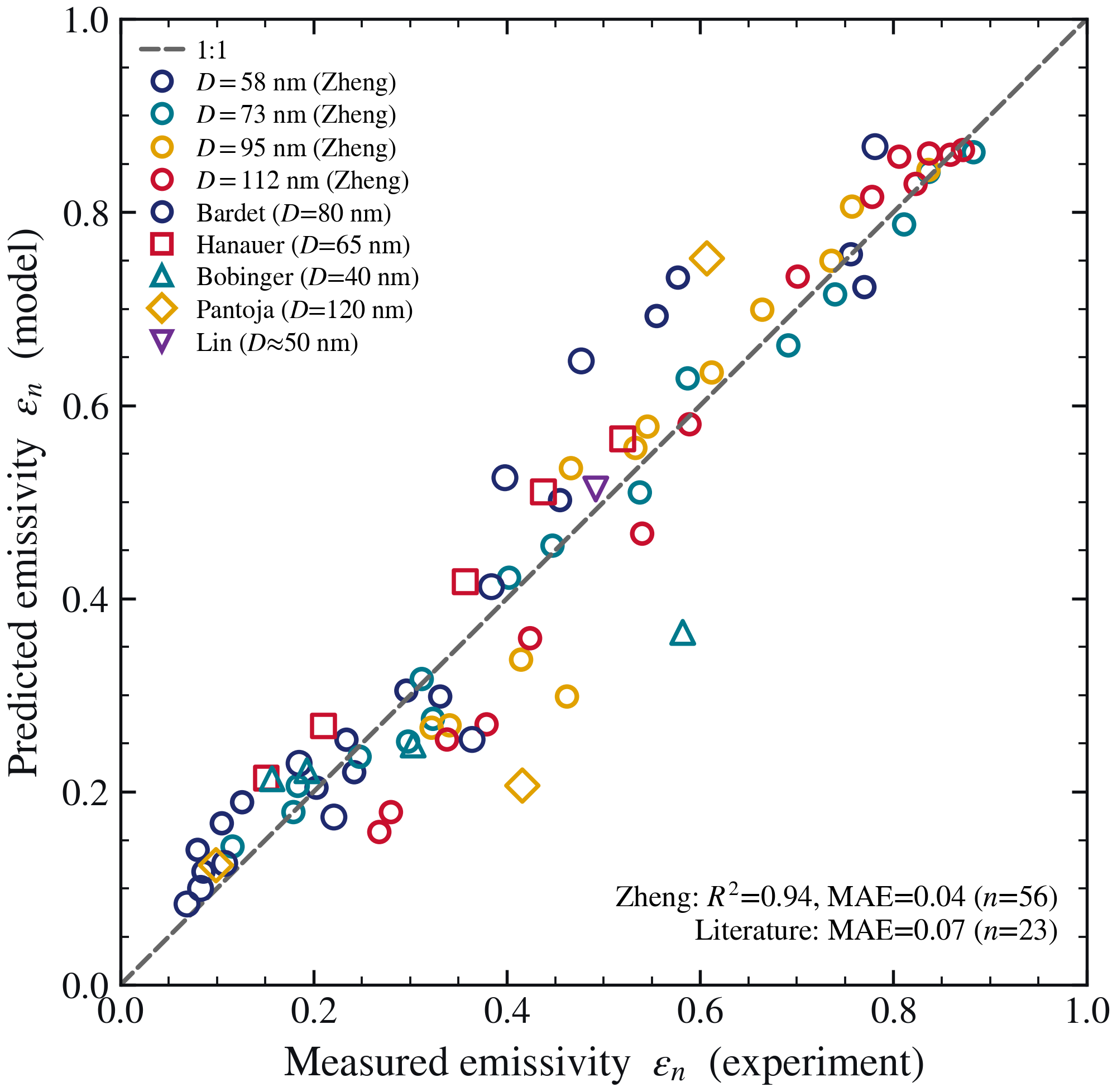}
\caption{Predicted (model) versus measured normal emissivity, all with the
\emph{same} global parameter set. Open circles coloured by diameter are the $56$
samples of Ref.~\cite{zhengFirst2026} (the calibration set, $D=58$--$\SI{112}{nm}$).
The remaining $23$ points are the external validation set---five further
independent laboratories, each drawn with its own symbol and colour:
Bardet~\cite{bardetOptimization2023a} ($D=\SI{80}{nm}$),
Hanauer~\cite{hanauerTransparent2021d} ($D=\SI{65}{nm}$),
Bobinger~\cite{bobingerInfrared2017a} ($D=\SI{40}{nm}$),
Pantoja~\cite{pantojaLow2017b} ($D=\SI{120}{nm}$), and
Lin~\cite{linDirect2019a} ($D=\SI{50}{nm}$). Their diameters
($40$--$\SI{120}{nm}$) \emph{bracket} the $58$--$\SI{112}{nm}$ calibration range on
both sides, so these are extrapolative predictions, not interpolations. The dashed
line is $1{:}1$; goodness-of-fit statistics are annotated in the panel.}
\label{fig:parity}
\end{figure}

Figure~\ref{fig:parity} shows the predicted versus measured normal emissivity for
all $56$ samples of Ref.~\cite{zhengFirst2026} ($R^2=0.94$, mean absolute error
$0.04$), together with the $23$ independent literature points at
$40$--$\SI{120}{nm}$ (mean absolute error $0.07$; Sec.~\ref{sec:external}). All
scatter about the $1{:}1$ line with a small overall bias (predicted minus measured,
$+0.019$ on the external set, so the model runs slightly high on average), but the
external agreement is looser than the in-sample agreement---a mean
absolute error $1.8$ times larger---and it is uneven between groups
(Table~\ref{tab:extmae}). Because the single $\alpha$ is fitted once across all
four diameters with no per-diameter freedom, the diameter ordering itself is a
consequence of the geometric law $g\propto D^2$, not of the fit.

\begin{table}[h]
\centering
\caption{Per-group agreement on the external validation set. Predictions use each
group's reported diameter and the global parameter set; no quantity is refitted.}
\label{tab:extmae}
\begin{tabular}{lccc}
\toprule
Group & $D$ (nm) & $n$ & MAE \\
\midrule
Bardet~\cite{bardetOptimization2023a}   & 80  & 10 & 0.066 \\
Hanauer~\cite{hanauerTransparent2021d}  & 65  & 5  & 0.060 \\
Bobinger~\cite{bobingerInfrared2017a}   & 40  & 4  & 0.089 \\
Pantoja~\cite{pantojaLow2017b}          & 120 & 3  & 0.127 \\
Lin~\cite{linDirect2019a}               & 50  & 1  & 0.021 \\
\midrule
Overall                                 & --- & 23 & 0.075 \\
\bottomrule
\end{tabular}
\end{table}

The largest group-averaged error is Pantoja's, whose $\SI{120}{nm}$ wires lie at the
top of the diameter range and whose three points span the widest inferred density
interval of any group. The two largest individual residuals are $0.22$ (Bobinger, at
their most transparent film, where the model underestimates) and $0.21$ (Pantoja, at
an intermediate density, in the same direction); in both cases the areal density is
not measured but inferred through the transmittance relation below, so part of the
discrepancy is carried by that step rather than by the infrared model. We report the
spread rather than the aggregate alone because the aggregate flatters the
comparison.

The result is insensitive to the literature material constants. Factor-of-two
changes in $\sigma_{\mathrm{dc}}$ or $\tau$ leave the global mean absolute error at
$0.042$, because $\alpha$ absorbs the rescaling (its optimum shifts to keep the
product $\alpha\,\sigma_{\mathrm{dc}}$ fixed)---the ``one optical degree of
freedom'' noted after Eq.~\eqref{eq:Rdc}. It is likewise robust to the silica
dispersion model: replacing our three-oscillator parametrization
[Eq.~\eqref{eq:sio2}] with an independent literature-style fused-silica model
(different oscillator strengths and dampings) changes the global mean absolute error
by less than $0.001$.

\subsection{Independent data sets and spectral shape}
\label{sec:external}

\begin{figure}[h]
\centering
\includegraphics[width=\linewidth]{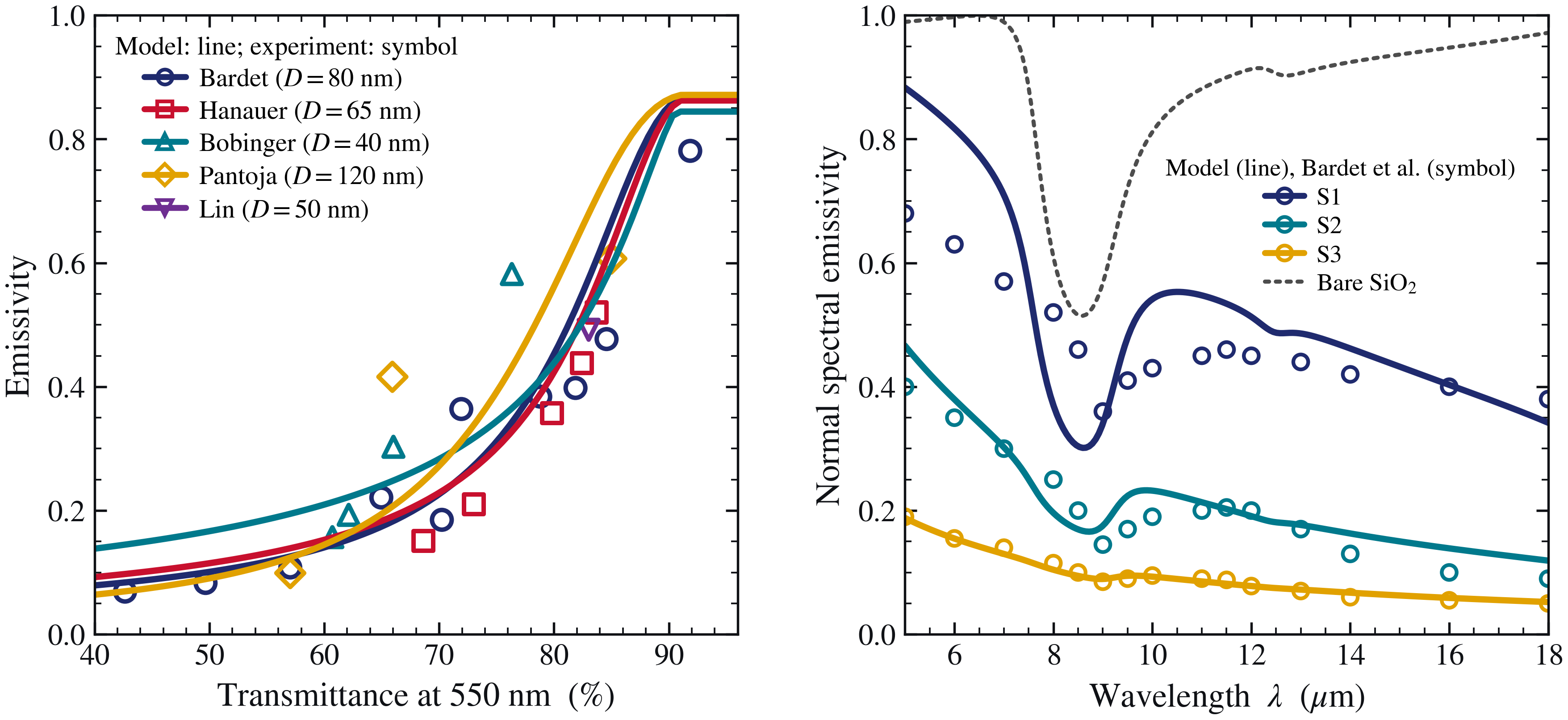}
\caption{Validation against independent data with identical parameters. (a)
Emissivity versus visible transmittance for five independent
groups~\cite{bardetOptimization2023a,hanauerTransparent2021d,bobingerInfrared2017a,pantojaLow2017b,linDirect2019a}:
each solid curve is the model evaluated across the full transmittance range at that
group's reported wire diameter ($40$--$\SI{120}{nm}$, spanning and extending beyond
the $58$--$\SI{112}{nm}$ set used to fix $\alpha$) and open symbols are their data. The
model predicts all groups (mean absolute error $0.075$ over $23$ points) from each
group's diameter alone; the reported wire lengths are not used, consistent with
length independence. The number of points per group is uneven ($10$, $5$, $4$, $3$,
and $1$ for Bardet, Hanauer, Bobinger, Pantoja, and Lin respectively): the Lin curve
is a model prediction across the full range constrained by a single measured point,
which tests the level at that point but not the trend. (b) Spectral emissivity of
three samples of Bardet \textit{et~al.}~\cite{bardetOptimization2023a} (symbols,
digitised from their published FTIR spectra; the digitisation is by eye and carries
an uncertainty of order $0.02$). The model curves use that work's reported diameter
($\SI{80}{nm}$) and areal densities ($84$, $186$,
$\SI{376}{\milli\gram\per\meter\squared}$) with the same global parameters and a
literature dispersive-silica substrate: nothing is fitted to the spectra, neither
level nor shape. The $\SI{9}{\micro\meter}$ dip and $\SI{11}{\micro\meter}$ recovery
are substrate reststrahlen features, visible in the bare-SiO$_2$ reference (dotted);
the mesh sets the level of each curve and the $1/\lambda$ decline beneath them.}
\label{fig:validation}
\end{figure}

Fixed entirely on the Ref.~\cite{zhengFirst2026} data, the model extends with
\emph{identical} parameters to data it has never seen (Fig.~\ref{fig:validation}).

\paragraph{Emissivity versus visible transmittance.}
Five independent
groups~\cite{bardetOptimization2023a,hanauerTransparent2021d,bobingerInfrared2017a,pantojaLow2017b,linDirect2019a}
report $\varepsilon$ against the visible transmittance $T$ at wire diameters of
$40$--$\SI{120}{nm}$, spanning and extending beyond the $58$--$\SI{112}{nm}$ range
used to fix $\alpha$. The $23$ points are unevenly distributed between them---Bardet
($10$ points, $D=\SI{80}{nm}$), Hanauer ($5$, $\SI{65}{nm}$), Bobinger ($4$,
$\SI{40}{nm}$), Pantoja ($3$, $\SI{120}{nm}$), and Lin ($1$,
$\SI{50}{nm}$)---so only the first four constrain a trend in $T$; the single Lin
point is a consistency check at one operating point, not a test of the shape of the
curve. Placing these points on a common axis uses only the calibrated
transmittance--coverage relation $T = T_0 - a\,(\mathrm{amd}/D)$
(Sec.~\ref{sec:geometry})~\cite{lagrange}; the infrared model is unchanged. The
model then predicts every group from its reported diameter alone---and, consistent
with length independence, without its wire length---to a mean absolute error of
$0.075$ over $23$ points [Fig.~\ref{fig:validation}(a)]. The emissivity itself is a
genuine prediction at these external diameters; the only Zheng-calibrated ingredient
is the relation that places points on the horizontal axis.

That ingredient is, however, not free of consequence, and it sets the accuracy
this test can claim. The relation is calibrated on one group's films and applied to
five others whose synthesis, substrate and transmittance convention differ, so it
carries the associated systematic: displacing $T_0$ by $\pm2\%$ moves the external
mean absolute error between $0.059$ and $0.101$. Its inversion also degrades at the
transparent end: Bardet's most transparent point ($T = 91.9\%$) maps to a slightly
negative areal density and is clipped to the sparsest physical value, so its
prediction is effectively the bare-glass ceiling and carries no information about
the mesh. It is retained in the quoted statistics for completeness---removing it
changes the overall mean absolute error only from $0.075$ to $0.074$---but it should
not be read as a test. The comparison as a whole should be taken as a test of level
and trend across diameters, not as a high-precision one.

\paragraph{Spectral emissivity.}
The tests above probe the level of $\varepsilon$; its spectral shape is a further,
independent check, and here nothing at all is adjusted. The three samples of Bardet
\textit{et~al.}~\cite{bardetOptimization2023a} have a reported wire diameter of
$\SI{80}{nm}$ and reported areal densities of $84$, $186$ and
$\SI{376}{\milli\gram\per\meter\squared}$; we feed those numbers to the model
together with the dispersive substrate of Eq.~\eqref{eq:sio2} and compare directly
with their measured $\varepsilon(\lambda)$ [Fig.~\ref{fig:validation}(b)], obtaining
a mean absolute error of $0.042$ over the three spectra with no free parameter---
neither the level nor the shape is tuned. The agreement is close for the two denser
films ($0.004$ and $0.029$) and looser for the sparsest, S1 ($0.093$), where the
model overshoots below $\SI{7}{\micro\meter}$ and again above $\SI{9.5}{\micro\meter}$
while cutting too deeply through the reststrahlen band itself---the sparsest network
is also the one whose gap is largest and whose response is closest to the bare
substrate, so it is the most sensitive to the substrate parametrization. The
band-integrated values that follow, $0.45$, $0.19$ and $0.08$, likewise sit close to
the $0.46$, $0.22$ and $0.09$ that work reports from an independent
(infrared-camera) measurement of the same three samples.

Of the spectral structure, the $\SI{9}{\micro\meter}$ dip and the
$\SI{11}{\micro\meter}$ recovery are reststrahlen features of the silica substrate:
they are already present in the bare-glass curve plotted alongside, and any model
carrying a correct substrate would show them. The AgNW network is responsible for
the overall level of each spectrum and for the soft $X_L\propto1/\lambda$ decline
on which those bands are superposed, and it is on those that the agreement is
informative.

\subsection{Length independence of the optics}

\begin{figure}[h]
\centering
\includegraphics[width=0.62\linewidth]{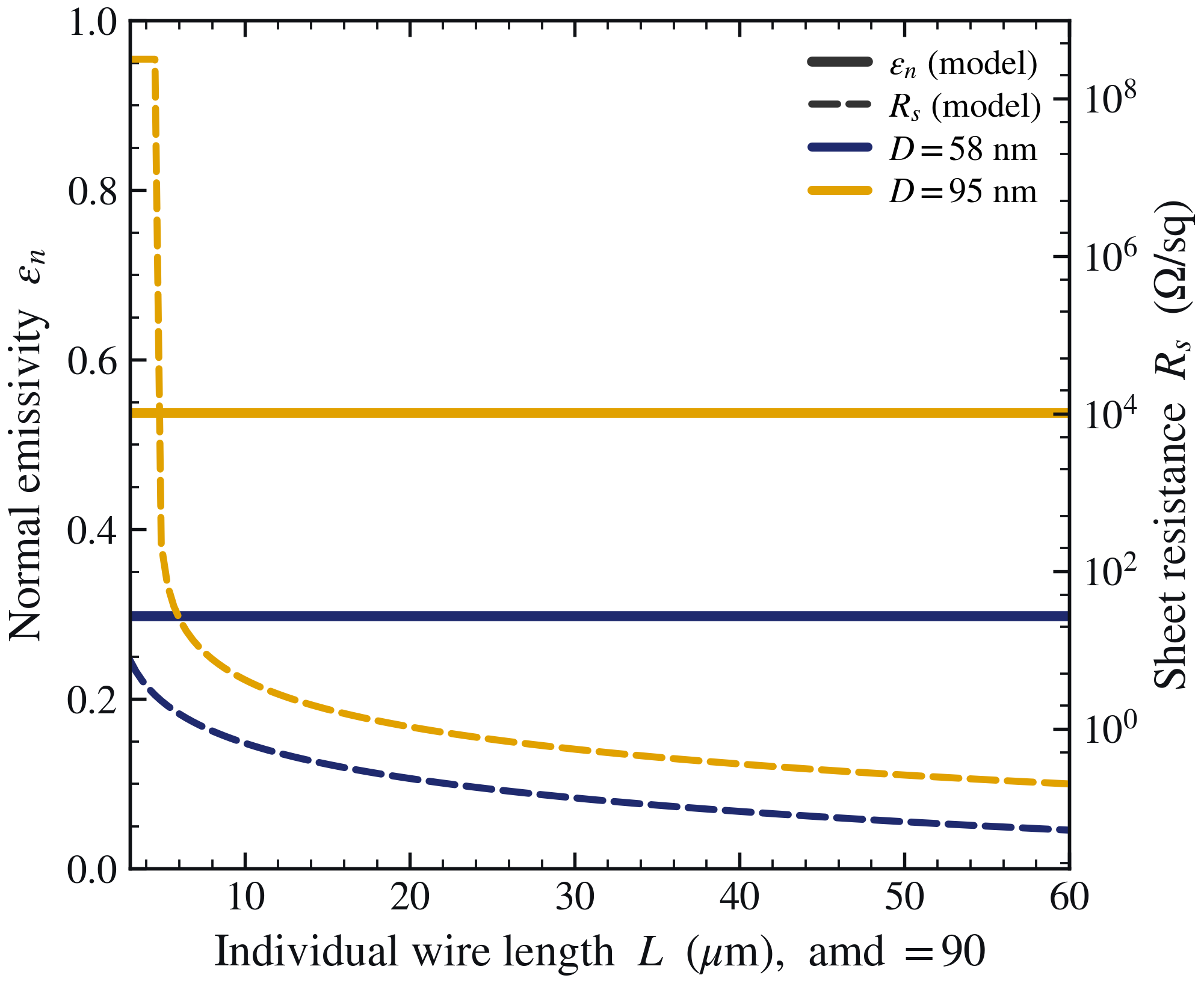}
\caption{Model normal emissivity (solid, left axis) is exactly independent of the
individual wire length $L$ at fixed areal density, whereas the
percolation-controlled dc sheet resistance (dashed, right axis) diverges as $L$ is
reduced toward the threshold.}
\label{fig:length}
\end{figure}

Because the gap [Eq.~\eqref{eq:gaplaw}] depends only on the areal density and the
diameter, and not on how the total length $\Lambda$ is partitioned into individual
wires, every optical quantity built on it is exactly independent of the wire length
$L$. Figure~\ref{fig:length} contrasts this with the dc sheet resistance from stick
percolation, $R_s \propto 1/(N - N_c)$ with $N_c \propto
L^{-2}$~\cite{liFinitesize2009}, which diverges as $L$ is shortened toward the
percolation threshold at fixed areal density. The optics is flat in $L$; the
transport is not. This is the cleanest statement that the infrared response is local
(gap-controlled) rather than topological (connectivity-controlled).

\subsection{Angular optical threshold}

\begin{figure}[h]
\centering
\includegraphics[width=0.68\linewidth]{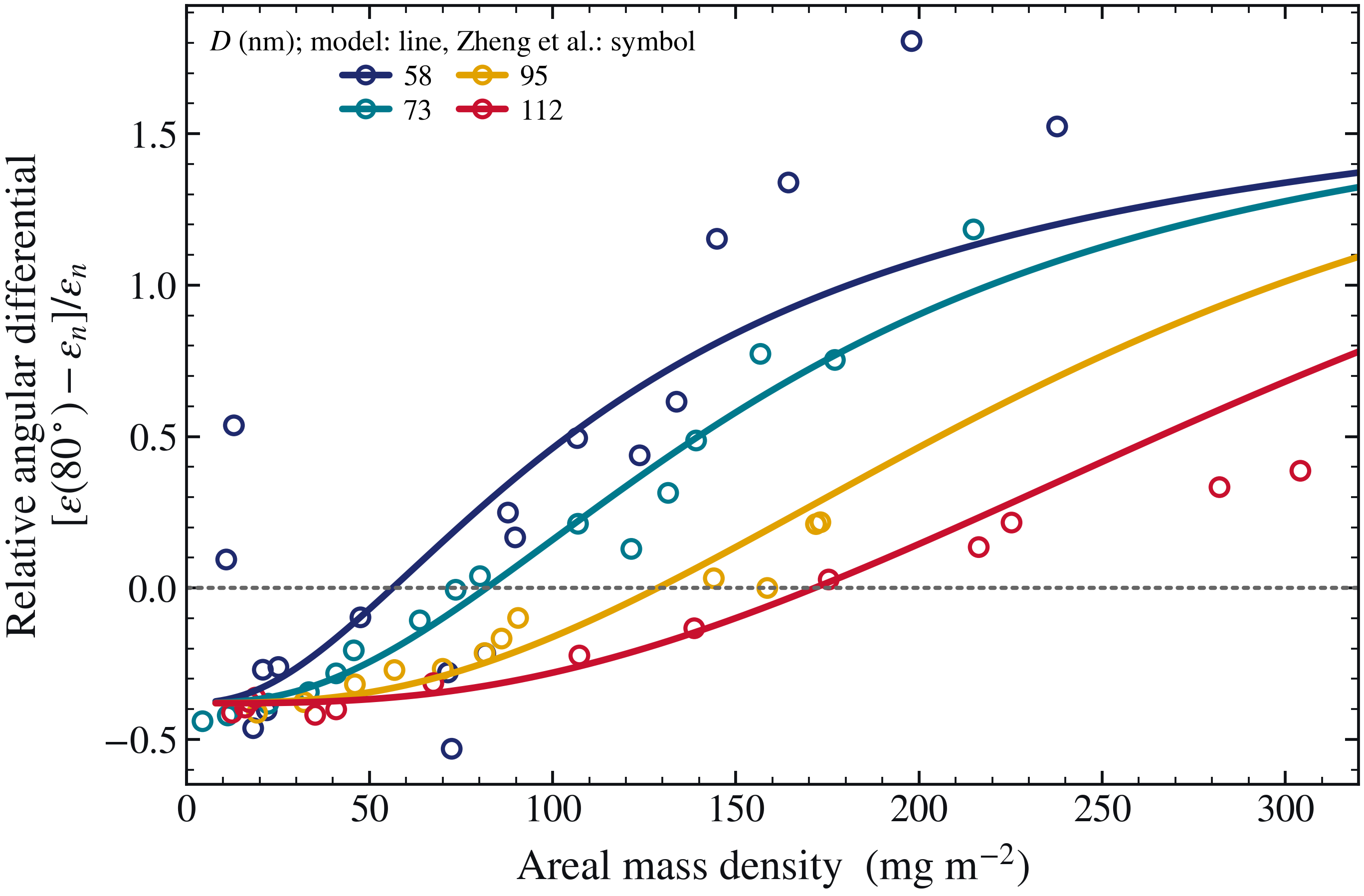}
\caption{Relative angular differential
$[\varepsilon(80^\circ)-\varepsilon_n]/\varepsilon_n$ versus areal mass density
(lines, model; open circles, data of Ref.~\cite{zhengFirst2026}). The zero crossing
marks the dielectric-to-metallic optical threshold and shifts to higher density for
thicker wires, as predicted by $g\propto D^2$.}
\label{fig:threshold}
\end{figure}

The directional emissivity changes character as the network densifies. We quantify
this with the relative angular differential
$[\varepsilon(80^\circ)-\varepsilon_n]/\varepsilon_n$, reported directly in the data
set. For sparse networks the weakly conducting sheet reflects like the bare
dielectric, so the emissivity falls toward grazing and the differential is
\emph{negative} (dielectric-like); for dense networks the metallic grazing-angle
peak makes it \emph{positive}. The model reproduces the sign change and predicts
that the zero crossing shifts to higher areal density for thicker wires
(Fig.~\ref{fig:threshold})---a direct consequence of the gap law $g\propto D^2$,
since a thicker-wire network needs a larger areal density to reach the same gap and
hence the same optical regime. The threshold is thus another manifestation of the
single control variable $g$, not an independent phenomenon.

\subsection{Comparison with the Hanauer semi-empirical model}

\begin{figure}[h]
\centering
\includegraphics[width=0.8\linewidth]{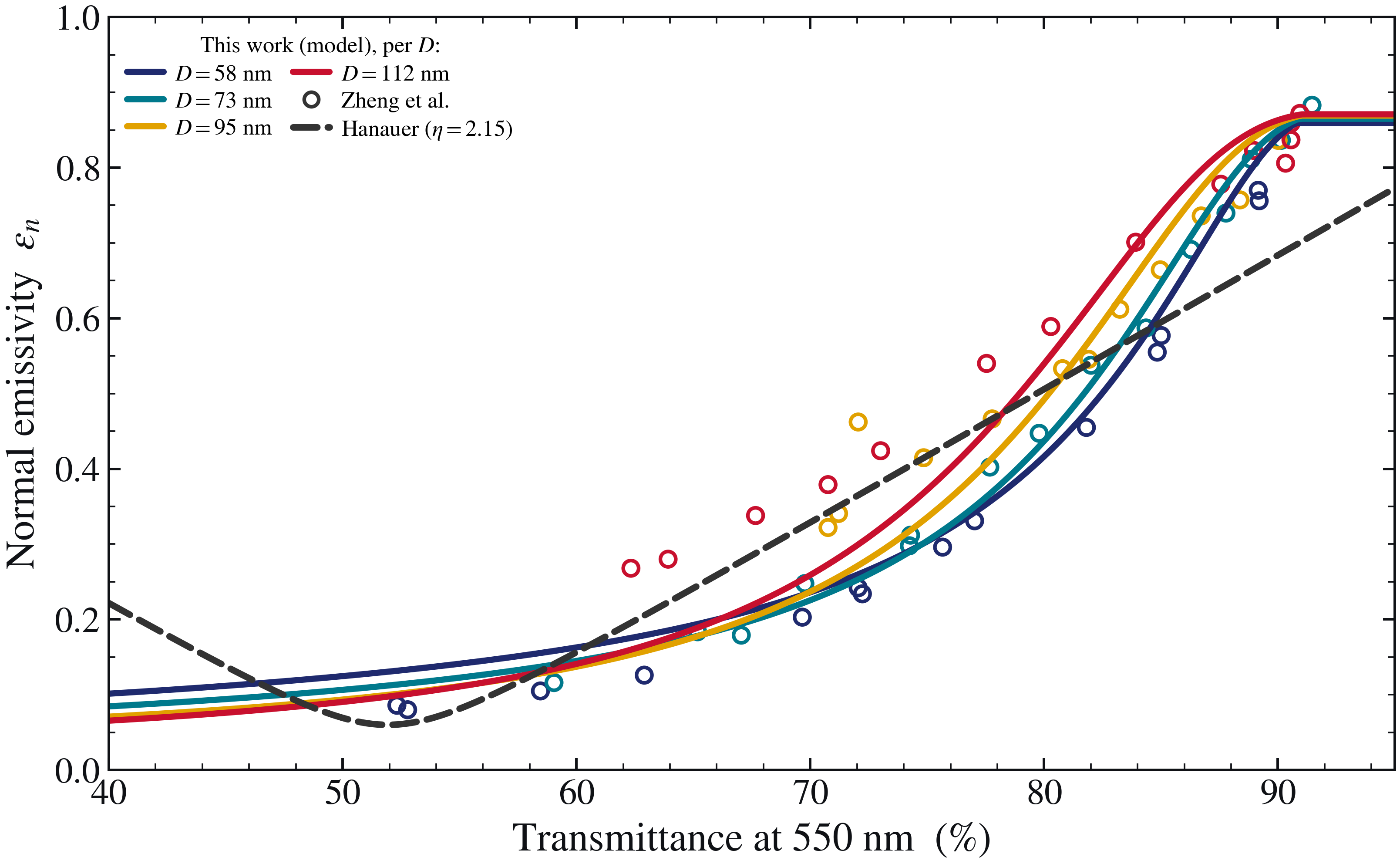}
\caption{Hanauer semi-empirical model [Eq.~\eqref{eq:hanauerTransparent2021d}] with a
single $\eta=2.15$ (dashed) against the present model evaluated per diameter (solid
lines) and the emissivity--transmittance data of Ref.~\cite{zhengFirst2026} (open
symbols, coloured by diameter). A single $\eta$ cannot follow the diameter spread
the present model captures, and it turns over unphysically toward low $T_r$
($\varepsilon\to0.93$ as $T_r\to0$ with these values, and $\to1.06$ with the
parameters published in Ref.~\cite{hanauerTransparent2021d}; in neither case is
there a metallic limit).}
\label{fig:hanauerTransparent2021d}
\end{figure}

The only prior attempt to describe $\varepsilon$ versus network density is the
semi-empirical expression of Hanauer \textit{et~al.}~\cite{hanauerTransparent2021d},
\begin{equation}
E_m = \sqrt{E_{m,\mathrm{Ag}}^2 +
\Big[E_{m,\mathrm{Ag}} + E_{m,\mathrm{sub}}(1-\eta)
+ T_r\big(\eta\,E_{m,\mathrm{sub}} - E_{m,\mathrm{Ag}}\big)\Big]^2},
\label{eq:hanauerTransparent2021d}
\end{equation}
with $T_r$ the visible transmittance, $E_{m,\mathrm{Ag}}$ and $E_{m,\mathrm{sub}}$
the emissivities of bulk silver and the bare substrate, and $\eta$ a single
empirical parameter which, by their construction, must exceed unity. For their own
series they report $E_{m,\mathrm{Ag}}=0.04$, a measured $E_{m,\mathrm{sub}}=0.94$,
and $\eta=2.17$. To compare on our data on the same footing as our own model, we fix
$E_{m,\mathrm{Ag}}=0.06$ and $E_{m,\mathrm{sub}}=0.86$ to the metallic floor and
bare-glass values of our Fresnel model and fit $\eta$ by the same criterion used for
$\alpha$; the optimum over the whole data set is then $\eta\approx2.15$
(Fig.~\ref{fig:hanauerTransparent2021d}).

The comparison exposes two failures. First, a single $\eta$ is one global curve and
cannot reproduce the diameter dependence: our model, with one fixed and physically
interpreted parameter, predicts each diameter separately (mean absolute error
$0.04$, versus $0.08$ for the single-$\eta$ Hanauer curve), and refitting $\eta$ per
diameter would require values from $\eta\approx1.9$ ($\SI{112}{nm}$) to $2.4$
($\SI{58}{nm}$) with no rule relating $\eta$ to $D$ or the microstructure. Second,
the required $\eta>1$ is \emph{unphysical} in the opaque limit, and this is already
true of the published parametrization: with their own $E_{m,\mathrm{Ag}}=0.04$,
$E_{m,\mathrm{sub}}=0.94$ and $\eta=2.17$, Eq.~\eqref{eq:hanauerTransparent2021d}
gives $E_m\!\to\!1.06$ as $T_r\!\to\!0$---an emissivity above unity, which no surface
can have. Our substitution of the Fresnel values is the milder case and still fails,
returning $E_m\!\to\!0.93$ (and $>1$ once $\eta$ is refitted for the thinner wires):
a dense silver mesh emitting like a near-blackbody rather than approaching a mirror,
with the curve turning up non-monotonically toward low $T_r$ where there are no data.
The Hanauer expression is a useful interpolation over the transmittance range where
it was calibrated, but is neither predictive across diameters nor physical outside
it.

\section{Methodology: the role of generative AI}
\label{sec:ai}

We record here how this work was produced, both for transparency and because the
process is itself a notable illustration of AI-assisted physical modelling. The
central model was originated by a generative AI assistant (Anthropic Claude, Opus 4.8
model), with
the authors framing the problem, supplying data and literature, and verifying
every step.

The authors provided the AI with a formally structured statement of the problem: the
physical context (thermal-IR emissivity of AgNW transparent electrodes), the set of
robust but unexplained experimental regularities [observations (i)--(v) of the main
text], the prior modelling attempts and why they fail (the Drude/Hagen--Rubens
mirror estimate, percolation- and effective-medium-derived conductivities, and the
Hanauer semi-empirical fit), and direct access to the experimental data of
Ref.~\cite{zhengFirst2026}. The AI was not given the mechanism, the governing
equation, nor any candidate framework.

From this input the assistant carried the model through, in one continuous line of
reasoning, to a validated result:
\begin{enumerate}
\item it identified the relevant but non-obvious body of theory---the
transmission-line / inductive-grid description of metallic meshes
(Marcuvitz~\cite{marcuvitzWaveguide1986}, Ulrich~\cite{ulrichFarinfrared1967a})
developed for \emph{periodic} frequency-selective surfaces---as the natural
framework for a problem that had been treated only through transport and
percolation;
\item it made the key conceptual link of connecting that periodic-grid theory to a
\emph{disordered} nanowire network by recognising that the inter-wire gap plays the
role of the grid period, so that the mesh reactance
$X_L = Z_0(g/\lambda)\ln(2g/\pi D)$ [Eq.~\eqref{eq:XL}] carries over with $g$ the
stochastic-geometry gap of Eq.~\eqref{eq:gaplaw}---the step that unifies the
diameter, density, and length behaviour under one geometric variable;
\item it implemented the resulting sheet-admittance model
[Eq.~\eqref{eq:sigma}] in code, computed the Planck-band emissivity through the
conducting-sheet Fresnel relations, and fitted the single efficiency $\alpha$;
\item it validated the predictions against the experimental data---the diameter
fan-out, the universal collapse, the $R_s$ decoupling, the angular transition, and
the external data sets of Sec.~\ref{sec:external}---and drafted an initial account of
the findings in text and mathematics.
\end{enumerate}
The cross-domain identification in steps~1--2---bridging the metal-mesh /
frequency-selective-surface literature and the stochastic-geometry and transport
description of random networks, two fields the authors of the present work had been
treating separately---is the step we regard as most striking, and the one that made
the present model possible.

The authors' role was to pose the problem in the structured form above and then to
verify, direct, and refine: checking each derivation and numerical result against the
data, iterating the presentation for clarity and rigour, supplying the comparison
literature and the external data sets, and specifying the figures and the lines of
explanation to be developed. All physical reasoning was independently checked, and
the authors take full responsibility for the correctness of the content. We report
this workflow explicitly because the useful contribution of the AI here was not
language or bookkeeping but the physical hypothesis itself; consistent with
publisher policy, the AI is a tool and is not, and cannot be, an author.

\bibliography{EmissivityBib}